\documentclass[12pt]{article}
 \pdfoutput=1
\usepackage[utf8]{inputenc}
\usepackage[authordate,bibencoding=auto,strict,backend=biber,natbib]{biblatex-chicago}
\usepackage{graphicx,graphics,color,booktabs}
\usepackage{amsmath,amssymb, bm}
\usepackage{amsthm}
\usepackage{mathrsfs}
\usepackage{mathabx}
\usepackage{algorithm}
\usepackage{algpseudocode}
\usepackage{setspace}
\usepackage[dvipsnames]{xcolor}
\usepackage{multirow}
\usepackage{caption}
\usepackage[compact]{titlesec}
\usepackage{subcaption}
\usepackage{placeins}
\usepackage{listings}
\usepackage{color}
\usepackage[dvipsnames]{xcolor}
\definecolor{dkgreen}{rgb}{0,0.6,0}
\definecolor{gray}{rgb}{0.5,0.5,0.5}
\definecolor{mauve}{rgb}{0.58,0,0.82}

\usepackage[margin=1.0 in, top = 1.0in]{geometry}

\newcommand{\DS}{\displaystyle}

\newcommand{\cD}{\mathcal{D}}

\newcommand{\cX}{\mathcal{X}}
\newcommand{\cY}{\mathcal{Y}}

\newcommand{\cW}{\mathcal{W}}

\newcommand{\R}{{\rm I}\kern-0.18em{\rm R}}
\newcommand{\h}{{\rm I}\kern-0.18em{\rm H}}
\newcommand{\K}{{\rm I}\kern-0.18em{\rm K}}
\newcommand{\p}{{\rm I}\kern-0.18em{\rm P}}
\newcommand{\E}{{\rm I}\kern-0.18em{\rm E}}
\newcommand{\Z}{{\rm Z}\kern-0.18em{\rm Z}}
\newcommand{\1}{{\rm 1}\kern-0.24em{\rm I}}
\newcommand{\N}{{\rm I}\kern-0.18em{\rm N}}

\newcommand{\argmin}{\mathop{\mathrm{argmin}}}

\newcommand{\var}{\mathrm{var}}
\newcommand{\cov}{\mathrm{cov}}

\newcommand{\mydegree}{$^{\circ}$}

\def\@begintheorem#1#2{\trivlist \item[\hskip \labelsep{\bf #1\ #2.}]\sl}
\def\@opargbegintheorem#1#2#3{\trivlist
      \item[\hskip \labelsep{\bf #1\ #2\ (#3).}]\sl}

\newtheorem{theorem}{Theorem}[section]

\newtheorem{lemma}[theorem]{Lemma}
\newtheorem{assumption}{Assumption}

\usepackage{hyperref}
\hypersetup{colorlinks=true,linkcolor=blue,citecolor=blue}
\usepackage{cleveref}
\graphicspath{ {images/} }

\title{\Large \vspace{-2cm} Estimating Continuous Treatment Effects in Panel Data using Machine Learning with a Climate Application 
}

\vspace{-3cm}

\author{%
  Sylvia Klosin\thanks{Department of Agricultural and Resource Economics,
    University of California, Davis. Email: \texttt{sklosin@ucdavis.edu}}%
  \and
  Max Vilgalys\thanks{New York City Mayor's Office of Management and Budget.
    Email: \texttt{vilgalysm@omb.nyc.gov}}%
  \and
  \thanks{Acknowledgements: We thank Isaiah Andrews, Victor Chernozhukov, Dalia Ghanem, Vitor Hadad, Chris Hansen, Namrata Kala, Jing Li, Anna Mikusheva, Whitney Newey, Dev Patel, Felix Pretis, Jonathan Roth, Liyang Sun, Jeff Wooldridge, and participants of the MIT Econometrics Lunch Seminar, the 2023 Africa Meeting of the Econometric Society session on Climate Econometrics, and the Harvard Climate Economics Pipeline Workshop for helpful comments. Klosin acknowledges support from the Jerry A. Hausman Graduate Dissertation Fellowship and the NSF Graduate Research Fellowship.}%
}
\date{ }
\begin{document}

\newtheorem{example}{Example}
\maketitle

\begin{abstract}

Economists often estimate continuous-treatment effects in panel data using linear two-way fixed effects models (TWFE). When the treatment–outcome relationship is nonlinear, TWFE is misspecified and potentially biased for the average partial derivative (APD). We develop an automatic double/de-biased machine-learning (ADML) estimator that is consistent for the population APD while allowing additive unit fixed effects, nonlinearities, and high-dimensional heterogeneity. We prove asymptotic normality and add two refinements—optimization-based de-biasing and analytic derivatives—that reduce bias and remove numerical approximation error. Simulations show that the proposed method outperforms high-order polynomial OLS and standard ML estimators. Our estimator leads to significantly larger (by 50\%), but just as precise, estimates of the effect of extreme heat on corn yield in comparison to standard linear models.   


\end{abstract}

\newpage

\section{Introduction} \label{sec::introduction}

Estimating the effect of a continuous treatment in panel data is central to applied economics:  key treatments such as distance, market access, prices, pollution concentrations, temperature, and tax rates are all continuous.\footnote{See, for example, \citet{saez2001_using_elasticities, autor_dorn_hanson_2013_china_syndrome, DonaldsonHornbeck2016}.} The workhorse model is the linear two-way fixed effects model (TWFE), which combines additive unit fixed effects with a linear specification in the treatment and covariates; when the true function is nonlinear, this leads to large specification error \citep{hastie2009elements}. A growing body of evidence shows strong nonlinearities in many economic relationships— e.g., weather and economic outputs, pollution and health, transit access and car ownership, and physical activity and health—underscoring the need for flexible methods.\footnote{See, for example, \citet{chay_greenstone_2003_infant_mortality, currie_neidell_2005_air_pollution_infant_health, Burke2015, aune2015_physical_activity_t2d}.}

One way to add flexibility into TWFE models is to use machine learning (ML).\footnote{In practice, applied work often adds flexibility by binning the treatment or by adding high-order polynomials and interactions. Binning is ad hoc: results hinge on researcher-chosen cutoffs and can induce bias and inefficiency. High-order polynomials can reduce bias but typically inflate standard errors, making estimates noisy. Our regularized ML approach preserves the needed nonlinear flexibility while maintaining small standard errors.} However, existing approaches that combine ML with panel data typically modify or restrict the structure of fixed-effects. This is problematic, since applied researchers rely on additive unit and time fixed effects to control for unobserved heterogeneity in order to strengthen the credibility of treatment effect estimation. In this paper, we propose an estimator to address this gap by preserving unrestricted additive fixed effects while incorporating ML to flexibly capture nonlinearities and heterogeneous treatment effects in a data-driven way. In doing so, it generalizes the linear TWFE framework and improves continuous treatment effect estimation by reducing specification bias while maintaining precision.

This paper makes three contributions. First, we develop a new automatic double machine-learning (ADML) estimator for panel data that is consistent for the population average partial derivative (APD)\footnote{An R package is available via 
\texttt{remotes::install\_github("MartinSpindler/hdm", ref = "p-refactor-autodml-pd")}.} and prove asymptotic normality.  Second, we introduce two computational refinements—optimization-based de-biasing (ten-fold bias reduction in simulations) and analytic APD derivatives—that improve accuracy and stability. Our third contribution is to demonstrate the method's practical relevance. In an application to U.S. corn yields under extreme heat, we find elasticities about 50\% larger than piecewise-linear benchmarks with maintained precision, implying materially higher damages.

Our first contribution is a novel ADML estimator for panel models with unrestricted additive fixed effects and high-dimensional treatment heterogeneity. We impose additive unit and time fixed effects, while leaving the remaining component of the model to be flexibly learned by modern machine-learning methods. Our semiparametric model is $Y_{i,t} = a_i + \gamma_{0,t}(D_{i,t}, X_{i,t}) + \varepsilon_{i,t}$, $a_i$ are unit fixed effects, $D_{i,t}$ is treatment, and $X_{i,t}$ are time-varying covariates.\footnote{$X_{i,t}$ includes time fixed effects.} This flexibility allows us to control for a rich set of covariates and to model treatment effects that vary across time, space, and along covariates, as well as along the dose–response curve.\footnote{That is, we allow the marginal effect of treatment to depend on the treatment level itself.}  

To establish the properties of this estimator, we extend the ADML framework \citep{chernozhukov2022automatic} to first-differenced data, yielding a high-dimensional ML analogue of classic nonparametric first-difference estimators \citep{henderson2008nonparametric}.  ML regularization addresses the curse of dimensionality that limits classic nonparametric estimators, allowing us to model rich treatment effect heterogeneity in a data-dependent manner. Consistency of our estimator requires approximate sparsity of the regression function, but we leave the unit fixed effects unrestricted—an assumption well suited to applied work where unobserved heterogeneity may affect every unit, yet only a subset of covariates and interactions are relevant. We prove asymptotic linearity for estimators based on \textit{differences} of high-dimensional functionals rather than just levels. This requires new theory: we establish mean-square continuity for average derivatives of differenced functionals (via a weak reverse Poincaré inequality). This in turn delivers root-$n$ consistency and asymptotic normality for the average partial derivative (APD), and the same framework extends to weighted average derivatives and coarse dose–response curves, giving applied researchers a menu of interpretable objects for continuous treatments.

Beyond our theoretical contributions, we introduce two empirical innovations to the ML de-biasing literature: (i) an optimization-based method for estimating the de-biasing component, and (ii) an analytical approach for calculating the APD. For the de-biasing step, we solve the (weakly) convex minimization problem with standard optimization packages. Unlike iterative methods that depend on researcher choices (e.g., number of iterations, 
initialization, step size; \citealp{chernozhukov2022automatic, chernozhukov2022riesznet}), convex solvers deliver 
numerically precise optima, reducing researcher degrees of freedom and yielding up to an order-of-magnitude lower bias \citep{boyd2004convex}. For the APD, we replace numerical derivative approximations \citep[e.g.][]{chartrand2017numerical, van2020numerical} with a stable analytical formula: taking partial derivatives  of basis functions (e.g., in Lasso) and averaging them directly. This eliminates numerical error in derivative 
calculation and produces more stable estimates.

Lastly, we apply our estimator to study the effect of climate change on U.S. corn yields. We estimate the elasticity of yields with respect to 
marginal increases in extreme heat, controlling for beneficial heat and precipitation. Fixed effects are crucial in this setting to isolate weather shocks \citep{Deschenes2007}, and nonlinearities in the 
temperature–yield relationship are well established \citep{Schlenker2009, PretisHendry2013}. Compared to the piecewise linear model of \citet{Burke2016}, our flexible approach produces estimates about 50\% larger, yet equally precise. This implies an additional \$5.4 billion in annual damages by 2050 under median climate scenarios. Finally, our estimated dose–response curve shows that damages from extreme heat are attenuated in counties already exposed to more extreme heat.

ML estimators adapted for panel data are still relatively scarce. We provide an estimator that simultaneously accommodates additive unit fixed effects and high-dimensional treatment-effect heterogeneity. Existing panel ML estimators typically deliver one but not both. \citet{belloni2016inference} develop a DML procedure that allows arbitrary additive fixed effects but focus on binary treatments and assume homogeneous effects\footnote{For binary treatments, homogeneity requires that treatment effects are the same across units. For continuous treatments, however, homogeneity imposes an additional strong restriction: the marginal effect of an increase in treatment must be identical at every point along the treatment distribution. This rules out the nonlinear dose–response relationships that are frequently observed in applied work.} — an assumption that can induce bias when effects vary across units and over time. Other work allows heterogeneous treatment effects while restricting fixed effects via correlated random effects, which constrain how time-invariant unobservables may relate to covariates \citep{semenova2023inference, chernozhukov2022automatic}; this restriction is not required under classic fixed effects. \citet{semenova2023inference} use correlated random effects and only allow sparse fixed-effect deviations, which allows them to include lags and dynamic terms. By contrast, we do not model dynamics, which enables us to leave the fixed effects completely unrestricted. This is consistent with the static-panel framework used in the empirical literature. A further distinction from \citet{semenova2023inference} is that we use ADML rather than DML for bias correction. For continuous treatments, ADML is more stable than DML because it avoids inverting small estimated conditional treatment densities, also known as generalized propensity scores \citep{klosin2021automatic}.

The recent panel-data literature on biases arising from heterogeneous treatment effects across groups and time underscores the need to model rich heterogeneity. For binary treatments, there is a large literature that shows pooled linear regressions that ignore such heterogeneity can deliver non-convex averages or even wrong-signed estimates.\footnote{\citet{callaway2020difference, dechaisemartin2020twfe, sun2020estimating, goodman-bacon2021did}.} In the binary-treatment case, \citet{wooldridge2021two} shows that identification can be restored by saturating the regression with treatment $\times$ group $\times$ time interactions. For continuous treatments, however, the problem is fundamentally more severe: it is not possible to construct an estimator that fully controls for unobserved heterogeneity without running into identification failures \citep{hoderlein2012nonparametric}.\footnote{\citet{hoderlein2012nonparametric}'s nonparametric model is $Y_{i,t} = \zeta_t(D_{i,t}, X_{i,t}, a_i, \varepsilon_{i,t} )$.} \citet{hoderlein2012nonparametric} prove that in fully nonseparable panels, generalized differencing identifies only a local average derivative for ``stayers,'' i.e.\ units whose covariates do not change across periods. Applied researchers, however, are often interested in population-level objects as they can have greater policy relevance. To achieve point identification of the population average derivative while preserving flexibility, we allow treatment effect heterogeneity to depend on a high-dimensional vector of covariates while assuming that unit fixed effects are separable.\footnote{Our semiparametric model is $Y_{i,t} = \zeta_t(D_{i,t}, X_{i,t}, a_i, \varepsilon_{i,t} ) = a_i + \gamma_t(D_{i,t}, X_{i,t}) + \varepsilon_{i,t}$.} This strikes a balance between the infeasible general case and applied relevance, delivering interpretable estimands under minimal additional assumptions. Our model nests \citet{wooldridge2021two} as a special case. 

This paper is among the few that address continuous treatments in TWFE-style settings. 
Unlike \citet{CallawayGoodmanBaconSantAnna2024NBER}, who identify ATT-style causal responses under generalized parallel trends in a low-dimensional setting, and \citet{deChaisemartinDHaultfoeuilleEtAl2022arXiv, deChaisemartinDHaultfoeuilleVazquezBare2024AEAPP}, who rely on stayers or parametric structure, our estimator point-identifies the population APD 
in a high-dimensional panel framework by combining additive fixed effects and machine-learning under a strict exogeneity assumption.

The paper is structured in the following way. Section \ref{sec::estimation} sets up the framework of the paper,  introduces the parameter of interest, and presents our estimator. The simulation design and the Monte Carlo results are given in Section \ref{sec::simulations}. Section \ref{sec::application} covers our application and provides descriptions of our data, and our results. Section \ref{sec::conclusion} concludes.

\section{Estimation} \label{sec::estimation}

\subsection{Notation and Definitions} \label{sec::notation}

We work in a panel data setting with $n$ individuals and $T$ time periods. 
As is often the case in economic data, we assume that $n$ is large, but  $T$ can be small.
We assume we have independent and identically distributed data $(W_1, \cdots, W_n)$ where  $W_i = \{(X_{i,t}, D_{i,t}, Y_{i,t})\}_{t=1}^T$ are copies of a random variable $W$ with support $\{ \cW = \cX \times \cD \times \cY\}_{t=1}^T$, with a cumulative distribution function (cdf) $F_{YDX}(Y,D,X)$. We use capital letters to denote random variables and lowercase letters to denote their possible values. For each unit in a population, $X_{i,t} \in \R^{h}$ denotes a vector of covariates, with $h$ potentially large, and $D_{i,t} \in \R$ denotes the treatment variable. We use $\E$ to denote the population expectation, and borrowing notation from the empirical process literature, we use $\E_n$ to denote the sample average so $\E_n(X_{i,t}) =\frac{1}{n T} \sum_{i = 1}^{n} \sum_{t =1}^{T} X_{i,t}$. 

For a given variable $X$, we use the notation $\Delta {X}_{i,t} :=  X_{i,t} - X_{i,t-1}$ for the first-difference transformation. For the first-difference transformation of a function $f$ of a variable $X$, we apply the function $f$ before taking the difference: $\Delta f(X_{i,t}) :=  f(X_{i,t}) - f(X_{i,t-1})$. Note that after applying a first-difference transformation, we remove the first time period. The empirical expectation in first-differenced data is therefore: $\E_n(X_{i,t}) =\frac{1}{n (T-1)} \sum_{i = 1}^{n} \sum_{t =2}^{T} X_{i,t}$. 

Define $|\cdot|_1$ as the $\ell_1$ norm; that is, $|\beta|_1 = \sum_{j=1}^p|\beta_j|$ where $\beta_j$ is the $j^\text{th}$ component of $\beta$ and $p$ is the length of $\beta$.

\subsection{Model} \label{sec::param of interest}

Our parameter of interest is the average continuous treatment effect $\tau_0$. To introduce the object we use potential outcome notation following \citet{colangelo2020double}. For every value of the continuous treatment $D_{i,t}$ we have a corresponding potential outcome $Y_{i,t}(D_{i,t})$.  
 
\begin{equation}
\label{eq:APD}
    \tau_0 = \E\bigg[\frac{\partial Y_{i,t}(D_{i,t})}{\partial D_{i,t}}\bigg].
\end{equation}

This $\tau_0$ is the effect of increasing the continuous treatment $D_{i,t}$ an infinitesimal amount averaging over all values of $D_{i,t}$ and $X_{i,t}$. This object is also known as an average derivative. When $Y_{i,t}$ is in log scale, this parameter captures the elasticity of $Y_{i,t}$ with respect to a marginal change in $D_{i,t}$. For example, in the case of our empirical example, $Y_{i,t}$ is log corn yields for US county $i$ in year $t$ and $D_{i,t}$ is exposure to extreme heat, and our $\tau_0$ is the elasticity averaging over all counties and years. 

To make estimating $\tau_0$ tractable, we impose some structure on the potential outcomes. In our paper we impose that:
\begin{equation}
\label{eqn:basic_regression}
Y_{i,t}(d_{i,t}) = a_i + \gamma_{0,t}(d_{i,t}, X_{i,t}) + \varepsilon_{i,t} \quad \quad E[\varepsilon_{i,t}| a_i, X_{i,1}, \cdots, X_{i,T}, D_{i,1}, \cdots, D_{i,T} ] = 0.
\end{equation}
In Equation \eqref{eqn:basic_regression} $a_i$ represents individual fixed effects, and $\gamma_{0,t}$ is a flexible high-dimensional function of treatment and covariates. The covariates $X_{i,t}$ can include time-varying covariates or interactions between time invariant variables, like group dummies which indicate which county a unit is in, and time-varying variables like treatment $D_{i,t}$. We have the subscript $t$ on the $\gamma_{0,t}$ function to emphasize that our covariates $X_{i,t}$ can include time dummies that themselves can also interact with other variables.  We do not assume that the fixed effects $a_i$ are sparse, but we do assume that $\gamma_{0,t}$ is approximately sparse and can be estimated consistently with Lasso. We assume strict exogeneity, an assumption  that is frequently used in applied work \citep{wooldridge2010econometric}. Our parameter of interest is, 

\begin{equation}
    \tau_0 = \E \bigg[\frac{\partial \gamma_{0,t}(D_{i,t}, X_{i,t})}{\partial D_{i,t}} \bigg], 
\end{equation}

which represents the average effect of a marginal increase in treatment, controlling for fixed effects. Identification follows directly from Equation \eqref{eqn:basic_regression} under the standard assumption of strict exogeneity. In Equation \eqref{eqn:basic_regression}
we impose structure on potential outcomes, namely that fixed effects are additive, as are the errors. Both of these assumptions are made when applied researchers run TWFE models. Our paper generalizes the commonly used TWFE model by estimating the regression function semiparametrically with machine learning, allowing for nonlinearities and high-dimensional interactions. We provide an example of what $\gamma_{0,t}$ looks like in a flexible regression specification below.   

\begin{example}(Treatment Heterogeneity in Group and Time)
\label{example_groups} 

 Equation \eqref{eqn:basic_regression} above presented a quite general model with its general function $\gamma_{t0}$. Below in Equation \eqref{eq:group_time_hetero} we present a concrete $\gamma_{t0}$ (that is nested within the model presented in Equation \eqref{eqn:basic_regression}) for a commonly run regression specification. This application has time-varying covariates enter linearly, and has rich heterogeneity in treatment effects across cohorts/groups and time.

\begin{equation}
\label{eq:group_time_hetero}
         Y_{i,t} = a_i + \underbrace{\beta Z_{i,t} + \sum_{s = 1}^T \sum_{g = 1}^G  \beta_{gs}(  1_{g} \times 1_{s}) +  \sum_{s = 1}^T \sum_{g = 1}^G  \tau_{gs}( D_{i,t} \times 1_{g} \times 1_{s})}_{\gamma_{0,t}(D_{i,t}, X_{i,t})}   +  \varepsilon_{i,t}.
\end{equation}
\end{example}

Here our covariates $X_{i,t} = (Z_{i,t}, 1_s, 1_g)$, where $Z_{i,t}$ are time-varying covariates, the $1_s$ are time dummies to allow time effects, and $1_{g}$ are the cohort/groups dummies. Cohort dummies indicate when a unit received a certain level of treatment.  We allow for interactions between covariates and treatment, creating a model with rich heterogeneity. Therefore, our estimand of interest $\tau_0$, the average partial derivative, is the population average of the effects of group/time treatment $\tau_{gs}$\footnote{Note here we focus on the average partial derivative as opposed weighted average partial derivative in the cross-sectional case in \cite{chernozhukov2022automatic}, which require exogenous weights (such as the normal distribution)}. While Equation \eqref{eq:group_time_hetero} can be estimated using OLS, the high dimensionality of the interaction terms leads to overfitting and a loss of efficiency. Our Lasso estimator mitigates this issue by imposing regularization that stabilizes estimation and preserves statistical power, as shown in the Monte Carlo results below.

Although our model imposes additive fixed effects, these are unit-specific and therefore do not restrict group/cohort-by-time heterogeneity. We present Example \eqref{example_groups} as a specific application of Equation \eqref{eqn:basic_regression} because it corresponds to a specification commonly employed in applied work and connects directly to the TWFE literature on negative weights. In particular, Equation \eqref{eq:group_time_hetero} corresponds to the main specification analyzed in \citet{wooldridge2021two}, expressed here in our notation. As emphasized by \citet{wooldridge2021two}, allowing for sufficiently rich heterogeneity in the outcome equation resolves the negative-weight problem that arises in conventional TWFE estimators.

\subsubsection{First-differences }

To deal with the fixed effect term $a_i$, we introduce a first-differenced version of Equation \eqref{eqn:basic_regression}. By taking a first-difference, we remove the time-invariant factor $a_i$ and consistently estimate $\gamma_{0,t}$. Once we remove the $a_i$ term, we can use cross-fitting in our final estimator while avoiding restricting the dependence structure within units. This is explained in detail in Section \ref{sec:Estimator}. 

We start by subtracting Equation \eqref{eqn:basic_regression} at time period $t-1$ from Equation \eqref{eqn:basic_regression} at time period $t.$
\begin{equation}
\label{eq:first_difference}
        Y_{i,t} - Y_{i,t-1} =  \gamma_{0,t}(D_{i,t}, X_{i,t}) - \gamma_{0, t-1}(D_{i,t-1}, X_{i,t-1}) + \varepsilon_{i,t} - \varepsilon_{i,t-1}.
\end{equation}
\begin{equation}
\label{eq:delta_first_difference}
        \Delta Y_{i,t}  =  \Delta \gamma_{0}(D_{i,t}, X_{i,t}) + \Delta \varepsilon_{i,t}. 
\end{equation}

Note for ease of notation, since the $\Delta \gamma_{0}$ is a function of multiple time periods, we drop the $t$ subscript. Note that the estimation target can be expressed equivalently in terms of the average derivative of $\Delta \gamma_{0}$, rather than $\gamma_{0,t}$:

\begin{align}
    \tau_0 
    &= \E\!\left[\frac{\partial \Delta \gamma_{0}(D_{i,t}, X_{i,t})}{\partial D_{i,t}}\right] \notag \\
    &= \E\!\left[\frac{\partial \gamma_{0,t}(D_{i,t},X_{i,t}) 
        - \partial \gamma_{0,t-1}(D_{i,t-1},X_{i,t-1})}{\partial D_{i,t}}\right] \notag \\
    &= \E\!\left[\frac{\partial \gamma_{0,t}(D_{i,t},X_{i,t})}{\partial D_{i,t}}\right].
    \label{eqn:average_derivative}
\end{align}

The last equality follows because given our structure $\gamma_{0,t-1}(D_{i,t-1}, X_{i,t-1})$ depends only on past treatment $D_{i,t-1}$, so its derivative with respect to $D_{i,t}$ is zero.

\subsection{Two-Stage Estimator}
We construct a de-biased estimator of the average derivative $\tau_0$ using two ML algorithms. 
First, we use Lasso to estimate the true regression function $\Delta \gamma_{0}$ by $\Delta \hat{\gamma}$ and calculate its average derivative (detailed in subsection \ref{sec:lasso details}). 
Estimating the regression function with ML can introduce overfitting and regularization bias.  To address these biases, we introduce a second ML problem to estimate a de-biasing term which we denote by $\alpha
_0$ and estimate with $\hat{\alpha}$ (detailed in subsection \ref{sec:Riesz Representer details}). Our estimator of $\tau_0$ combines $\Delta \hat{\gamma}$ and $\hat{\alpha}$ and is given at the end of this section in subsection \ref{sec:Estimator} in Equation \eqref{eq:estimator}.

\subsubsection{Regression Function and Its Derivative}
\label{sec:lasso details}

We use Lasso to estimate $\Delta \gamma_{0}$ by $\Delta \hat{\gamma}$, and use it to compute the derivative $\frac{\partial \Delta \hat{\gamma}(D_{i,t}, X_{i,t})}{\partial D_{i,t}}$. We assume that $\gamma_{0,t}$ is approximately sparse, which means that Lasso can estimate $\gamma_{0,t}$ well. This assumption is formalized in Appendix \ref{sec:Primitive ConditionsforAssumption2}. 

There are two general steps to find this derivative of the regression function 
\begin{enumerate}
    \item \textbf{Estimate $\Delta \hat{\gamma}$. }

\begin{enumerate}
    \item
    Transform the covariates $\{D_{i,t}, X_{i,t}\}$ using a set of basis functions\footnote{Basis functions are a set of functions that take arbitrary transformations of variables, and can be used to express a wide category of transformations such as polynomial expansion, splines, and radial basis functions.}  into a high-dimensional set of covariates. 
We define our $p \times 1$ dictionary of basis functions by $b(D_{i,t}, X_{i,t}) \in \R^p$. These basis functions can include any desired transformations of the covariates, such as polynomial terms or interactions between variables.\footnote{ We use polynomial basis functions of terms and interactions, although other approaches like kernel functions or splines could be used as long as the derivatives are bounded. For example, we set $b(D_{i,t}, X_{i,t})$ to be a third-order polynomial set of the covariates and interactions between $D$ and each covariate in $X$.}

\item Then let $\Delta b(D_{i,t}, X_{i,t}) := b(D_{i,t}, X_{i,t}) - b(D_{i,t-1}, X_{i,t -1})$. 
We set each function in the dictionary $b$ to have mean $0$ and variance $1$; Appendix \ref{sec:normalization} defines the standardization and describes how to use the standardized basis functions in the remaining estimation.
\item Find a vector of coefficients $\hat{\beta}$  for our dictionary such that $\Delta \hat{\gamma}(D_{i,t},X_{i,t}) := \Delta b(D_{i,t},X_{i,t})' \hat{\beta}$ is an approximately sparse linear approximation of $\Delta \gamma_{0}(D_{i,t},X_{i,t})$\footnote{ Formal assumptions needed for Lasso to approximate our function well are given in\eqref{sec:Primitive ConditionsforAssumption2}}.
We do so by solving the following Lasso problem:
\begin{equation}
    \hat{\beta} = \argmin_{\beta} \left\{\frac{1}{n (T - 1)} \sum_{i = 1}^{n} \sum_{t =2}^{T} (\Delta Y_{i,t} - \Delta b(D_{i,t}, X_{i,t})' \beta)^2 + r_L|\beta|_1 \right\}.
\end{equation}
This procedure depends on the regularization weight $r_L$, which we select by finding the value that minimizes test-set error in a cross-folds procedure. This procedure is described in Appendix \ref{sec:tuning alpha and gamma}. Researchers may also specify a weight vector at this stage if they are interested estimating a weighted Lasso. 
    
\end{enumerate}
    \item \textbf{Calculate the derivative.}

    We calculate the derivative analytically.
    Past DML approaches used numeric differentiation; we discuss this alternative in Appendix \ref{sec:numerical differentiation}. Using the analytical derivative, we do not introduce error from numerically approximating the derivative.  Our procedure uses the estimate of $\hat{\beta}_{\ell}$ from the previous step to compute the derivative of each function in our dictionary of basis functions. 
    
    \begin{enumerate}
        \item Construct the dictionary $b_D$, a $p \times 1$ dictionary of derivatives of each basis function in $b$.
        For each basis function $b^{j}$ for $j = 1, \dots, p$ in our dictionary of basis functions, define its derivative as follows: 
\begin{equation}
    b_D^{j}(D, X) = \frac{\partial b^{j}(D,X)}{\partial D}.
\end{equation}
\item Estimate the average derivative as: 
\begin{equation}
    \E_n\left[{\frac{\partial \Delta \hat{\gamma}(D_{i,t}, X_{i,t})}{\partial D_{i,t}}}\right] = \E_n[b_D(D_{i,t},X_{i,t})'\hat{\beta}].
\end{equation}
    \end{enumerate}
\end{enumerate}

\begin{example}

Consider a simple setting where $X_{i,t} \in  \R$, and where $ \gamma_{0,t}(D_{i,t}, X_{i,t}) = a_i +  {\beta}_1  D_{i,t} + {\beta}_6  D^2_{i,t} X_{i,t}^2$. We want to estimate the average derivative $ \E\left[{\frac{\partial {\gamma_{0,t}}(D_{i,t}, X_{i,t})}{\partial D_{i,t}}}\right]$. 
Our original basis function dictionary is $b(D_{i,t},X_{i,t}) = \{D_{i,t}, X_{i,t}, D_{i,t}X_{i,t}, D_{i,t}^2, X_{i,t}^2, D^2_{i,t} X_{i,t}^2 \}$. Our difference dictionary is $\Delta b(D_{i,t},X_{i,t}) = \{\Delta D_{i,t}, \Delta X_{i,t}, \Delta D_{i,t}X_{i,t}, \Delta D_{i,t}^2, \Delta X_{i,t}^2, \Delta D^2_{i,t} X_{i,t}^2 \}$

In step 1, we obtain an estimate $\hat{\beta} = (\hat{\beta_1}, \cdots, \hat{\beta_6})$ using Lasso using $\Delta b(D_{i,t},X_{i,t})$. There is one coefficient for every basis function. 
In step 2, we take the derivative of the orginal basis functions. 
Here, $b_D(D_{i,t},X_{i,t}) = \{1, 0, X_{i,t}, 2 D_{i,t}, 0, 2D_{i,t} X_{i,t}^2 \}$. 
The estimated average derivative is then: $\hat{\beta}_1 + \hat{\beta}_6 E_n[2D_{i,t}X_{i,t}^2]$.
\end{example}

\subsubsection{De-biasing term}
\label{sec:Riesz Representer details}

Our de-biasing term is based on the methods of \citet{chernozhukov2022automatic}, which uses the Riesz Representation theorem. The authors introduce the de-biasing term $\alpha_0$, which will satisfy Equation \eqref{eq:rr} for all functions $\gamma$: 

\begin{equation}
\label{eq:rr}
    \E\left[{\frac{\partial \Delta {\gamma}_{0}}{\partial D_{i,t}}}\right] = \E[\alpha_0 \gamma_{i,t}].
\end{equation}
This equality holds regardless of the function $\gamma$, so it is possible to estimate $\hat{\alpha}$ from data by plugging in various functions into this equality. Specifically, we use all $p$ functions inside the $p$ dimensional basis function dictionary $b$, as defined in the last section. We write our equality now using $p$ dimensional vectors. 
\begin{equation}
\E\left[ b
_D
 (D_{i,t}, X_{i,t})\right]  = \E[\alpha_0 \Delta b(D_{i,t}, X_{i,t})].  
\end{equation}
To construct an estimate $\hat{\alpha}$, we introduce an empirical version of the above equation.  
\begin{equation}
\label{eq:rr_emperical}
   \E_n\left[  b_D
 (D_{i,t}, X_{i,t})\right] = \E_n[\hat{\alpha} \Delta b(D_{i,t}, X_{i,t})].  
\end{equation} 
We assume that $\alpha_0$ has a sparse linear form: $\alpha_0(D_{i,t}, X_{i,t}, D_{i,t-1}, X_{i,t-1}) = \Delta b(D_{i,t}, X_{i,t})' \rho_0$. This assumption is formalized in Appendix \ref{sec:Primitive ConditionsforAssumption2}. Because $\alpha_0$ is a function of the $\Delta b$ dictionary, it is a function of current and past data values. Our estimate is $\hat{\alpha}(D_{i,t}, X_{i,t}, D_{i,t-1}, X_{i,t-1}) = \Delta b(D_{i,t}, X_{i,t})' \hat{\rho}$. 
Plugging the definition of $\hat{\alpha}$ into Equation \eqref{eq:rr_emperical}, we get:
\begin{equation}
\label{eq:rr_plugin}
    \E_n\left[  b_D
 (D_{i,t}, X_{i,t})\right] = \E_n[\Delta b(D_{i,t}, X_{i,t})' \hat{\rho}  \Delta b(D_{i,t}, X_{i,t})].  
\end{equation}
To solve for $\hat{\alpha}$, we find the parameter vector $\hat{\rho}$ that best makes the $p$ equalities in Equation \eqref{eq:rr_plugin} hold. Rearranging and adding a regularization term, we can find $\hat{\rho}$ by minimizing Equation \eqref{eq:squared_loss_simplified}. Additional steps are given in Appendix \ref{sec::origin alpha}. 
\begin{equation}
\label{eq:squared_loss_simplified}
    \hat{\rho} = \argmin_\rho \left\{ -2\E_n[b_D(D_{i,t}, X_{i,t})]\rho + \rho ' \E_n[\Delta b(D_{i,t}, X_{i,t})'\Delta b(D_{i,t}, X_{i,t})] \rho + r_\alpha|\rho|_1 \right\}. 
\end{equation}
Our final estimate of the de-biased term uses this estimate of $\hat{\rho}$ from Equation \eqref{eq:squared_loss_simplified}:
\begin{equation}
\label{eq:alpha_hat}
    \hat{\alpha}(D_{i,t}, X_{i,t}, D_{i,t-1}, X_{i,t-1}) = \Delta b(D_{i,t}, X_{i,t})' \hat{\rho}. 
\end{equation}
Additional details about this solution are given in Appendix \ref{sec::origin alpha}, including notation that matches the problem introduced by \citet{chernozhukov2022automatic}. 
\citet{chernozhukov2022automatic} provide an iterative procedure to solve for $\hat{\rho}$. 
We implement their procedure and also introduce an approach using an optimization package to solve the minimization problem in Equation \eqref{eq:squared_loss_simplified}.
Our optimization-based implementation guarantees that we find an optimal solution to this minimization problem (up to the level of numerical precision), as this function is weakly convex. 
The iterative approaches in the current literature are sensitive to researcher choices such as number of iterations, vector initialization, and step size, and may not converge to the true value of the parameter.
In simulation trials, we compare the performance of iterative and optimizer approaches for determining $\hat{\alpha}$. 

To illustrate the estimation procedure for a Riesz representer, we consider a simple example. 
\begin{example} For clarity, we focus on a one-dimensional, non-panel case.
  Consider the average derivative of a function $\gamma(D)$, and where the basis function is the identity (i.e. $b(D) = D$). 
Plugging the form of $b$ into Equation \ref{eq:squared_loss_simplified} and ignoring the regularization term for simplicity, we have:
\[
\hat{\rho} = \argmin_\rho \left\{ -2\rho + \rho ' \E_n[D^2] \rho \right\}. 
\]
Say  $D \sim \mathcal{N}(0,1)$, so that $\E_n[D^2] = 1$. 
Therefore, we have $\hat{\rho} = 1$. 

Our estimate of the Riesz representer is therefore $\hat{\alpha}(D) = D$. 
Reassuringly, this is the same as the Riesz representer derived using  integration by parts. 
After expanding the expectation,  $\DS \E\left[ \frac{\partial \gamma(D)}{\partial D}\right] = \E[D \gamma(D)]$.
That is, the true Riesz representer of the average derivative is $\alpha_0(D) = D$ when $D \sim \mathcal{N}(0,1)$. This also follows from Stein's Lemma. 
\end{example}

\subsubsection{De-Biased Estimator}
\label{sec:Estimator}

Our de-biased estimator combines estimates of the average derivative using $\hat{\gamma}$ and $\hat{\alpha}$ from above, and uses a cross-fitting approach to avoid overfitting bias. 

\paragraph{Cross-Fitting}
 
We use a cross-fitting approach to estimate both $\hat{\gamma}$ and $\hat{\alpha}$; this is a systematic approach to separate the datasets used to estimate functions and evaluate the average derivative. This mitigates overfitting by ensuring independence between training and evaluation samples \citep{hastie2009elements}. 
First, the researcher chooses the number of splits $L$ ($L = 5$ is commonly used). Then each unit's indices are randomly partitioned into the $L$ equally sized groups. We use $\ell$ to denote these groups, $\ell = 1, \dots, L$. Denote observations in group $\ell$ by $W_{\ell}$. Our functions $\hat{\gamma}_\ell$ and $\hat{\alpha}_\ell$ are trained using observations not in group $\ell$, then evaluated on observations in group $\ell$. 

To maintain the independence of data across folds, we keep all observations of a unit in the same fold. This allows for arbitrary dependence in the data within unit, which is important because observations within units are often correlated. Therefore, if observations of a unit were split across folds, the folds would no longer be independent. 

Because the individual fixed effect $a_i$ terms are unit specific and would therefore be found only in one fold, they cannot be estimated via our cross-fitting procedure. It was therefore critical to first-difference the data. This problem of cross-fitting while dealing with dependent data is an interesting problem for high-dimensional panel data; some papers deal with it by assuming weak dependence \citep{semenova2023inference}, and others risk overfitting and own observation bias by not cross-fitting at all. 

\paragraph{Score} For each observation, we define the de-biased score $\hat{\tau}_{\ell,i,t}$:
\begin{equation}
\label{eq:de-biased score cross-folds}
\hat{\tau}_{\ell; i,t} = \frac{\partial \Delta \hat{\gamma}_{\ell}(D_{i,t}, X_{i,t})}{\partial D_{i,t}} 
        +  \hat{\alpha_{\ell}}(D_{i,t}, X_{i,t}, D_{i,t-1}, X_{i, t- 1} )(\Delta Y_{i,t} - \Delta \hat{\gamma}_{\ell}(D_{i,t}, X_{i,t})). 
\end{equation}

We construct our estimate $\hat{\tau}$ as the average of the de-biased score from all folds. 

\paragraph{Estimator}
\begin{equation}
        \hat{\tau} = \frac{1}{n (T - 1)} \sum_{\ell=1}^{L} \sum_{i \in \ell} \sum_{t = 2}^{T} \hat{\tau}_{\ell; i,t}. \label{eq:estimator}
\end{equation}

This score defines a doubly-robust estimator, remaining unbiased even if either $\Delta \hat{\gamma}$ or $\hat{\alpha}$ is misspecified.

To compute the asymptotic variance of the estimator, we account for correlation of the average derivative within panel units. 
This clustered form of the variance is discussed in \citet{belloni2016inference}. Other ways of clustering errors for the DML setting are given in \citet{chiang2022multiway}. 
The asymptotic variance is: 

\begin{equation}
    \hat{V} = \frac{1}{n (T - 1)} \sum_{\ell=1}^L \sum_{i \in \ell}  \sum_{t =2}^T\sum_{t' =2}^T\hat{\psi}_{\ell; it}\hat{\psi}_{\ell; it'}. 
\end{equation}

where 
$\hat{\psi}_{ i,t} = \frac{\partial \Delta \hat{\gamma}}{\partial D_{i,t}} + \alpha(W_{i,t}) (\Delta y_{i,t} - \Delta \hat{\gamma}(D_{i,t}, X_{i,t})) - \hat{\tau}$.

Assumptions and the proof for asymptotic normality of our estimator are given in Appendix \ref{section:asymptotic_normality}.
Recall that we work with $(T - 1)$ time periods rather than $T$ because we removed one time period by first-differencing the data.

\subsubsection{Asymptotic Normality}

We leave proof, technical assumptions, and discussion of the formal asymptotic normality results to Appendix \ref{section:asymptotic_normality}, but state our main theoretical result here.  

\begin{theorem}(Asymptotic Normality) Under (i) the mild mean square conditions in Assumption~\eqref{assump:mild_mean_square}, 
(ii) the rate conditions on both the model and bias correction terms in Assumption~\eqref{assump:interaction_rate}, 
and (iii) Assumption~\eqref{assump:double_robust}, our estimator is asymptotically linear. 
Note that the number of time periods $T$ is fixed.

\label{theor:asymptotic_normality}

\begin{equation}
\sqrt{nT}(\hat{\tau} - \tau_{0}) \xrightarrow[]{d} \mathcal{N}(0, V), \quad \hat{V} \xrightarrow[]{p} V 
\end{equation}

Where 

\begin{equation}
\label{eqn:true_asymptotic_variance}
    V = \frac{1}{(T-1)}  \sum_{t =2}^T\sum_{t' =2}^T \E[\psi_{i,t}\psi_{it'}]
\end{equation}
and 
$\psi_{ i,t} = \frac{\partial \Delta \gamma_{0}}{\partial D_{i,t}} - \tau_0$.
\end{theorem}

\section{Simulations} \label{sec::simulations}

We assess the estimator's performance using Monte Carlo simulations. We design the data-generating process (DGP) to include nonlinearities, interactions, and correlations of treatment effects over time , ensuring that the setting is challenging and realistic.
We compare the performance of our DML estimator with three other models: (1) ordinary least squares (OLS) using the untransformed set of covariates, which we label ``OLS Linear''; (2) OLS using the polynomial expansion, which we label ``OLS Poly''; and (3) Lasso without a bias correction term, which we label ``Lasso''. When we implement our DML method, we calculate the bias correction using two different software approaches. We call our optimizer-based DML approach just ``DML''. We also adapted the iterative estimation procedure from \citet{chernozhukov2022automatic} for panel data, and label those results ``DML Iterative''. 
We estimate all models on 1,000 sample datasets constructed using our DGP. 
We then report the mean true value of the derivative, the mean of our estimates, the average bias of our estimates, and the mean squared error (MSE) between true and estimated values. 

\paragraph{DGP}: For each dataset, we draw a random sample with $N = 1000$ individuals, $T = 2$ time periods, and $h = 20$ original number of $X_{i,t}$ covariates. 
Our basis function transformation takes third-order polynomials of each variable, then adds interactions between each $\{D_{i,t}, X_{i,t}^{(j)}\}$ pair and their polynomials. We have a total of $p=244$ covariates after applying the basis function transformation.

We generate outcome variables according to the following function:
\begin{equation}
    Y_{i,t} = a_i + D_{i,t} + D_{i,t}^2 + D_{i,t}^3 + D_{i,t} X^{(1)}_{i,t}+ .1 \theta \mathbf{X}_{i,t}  + \varepsilon_{i,t}.
\end{equation}
To match real-world panel settings, we impose a correlation between $a_i, D_{i,t} $ and $\mathbf{X}_{i,t}$. 
We set $\theta$ so that the $j^\text{th}$ element is $\theta_j = 1/j^2$.
Fixed effects $a_i$, covariates $\mathbf{X}_{i,t}$, and random noise $\varepsilon_{i,t}$ are drawn from Gaussian distributions: $a_i \sim \mathcal{N}(1,1)$, $X^{(j)}_{i,t} \sim \mathcal{N}(a_i, 1) \ j = 1,\dots,h$, and $\varepsilon_{i,t} \sim \mathcal{N}(0,1)$, while the treatment is correlated with $\mathbf{X}_{i,t}$ but includes simulation draws from a Beta distribution:  $D_{i,t} \sim .1 \theta \mathbf{X}_{i,t} + Beta(1, 7) $

\begin{figure}[ht]
    \centering
    \includegraphics[width=\textwidth]{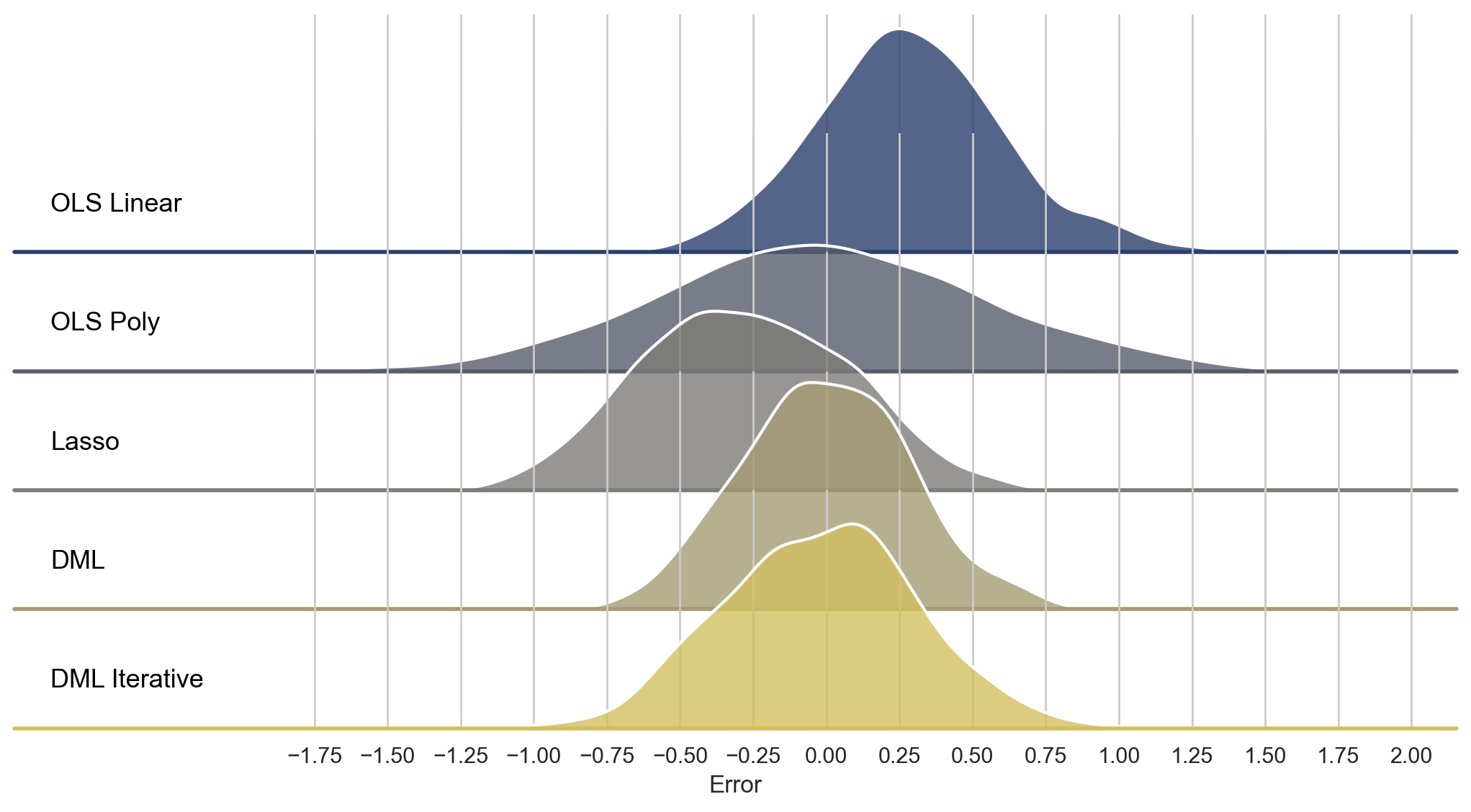}
    \caption{Visualization of distributions of error per simulation trial, for the five methods above. Error is defined as estimated value minus true value. }
    \label{fig:ridge plot of sims }
\end{figure}

We show the results of this simulation trial by summarizing the bias and MSE in Table \ref{tab:simulation results} and plot the distribution of the errors in Figure \ref{fig:ridge plot of sims }. In short, our DML estimator has lower bias than the very flexible OLS Poly and yet still maintains lower standard errors than OLS Linear.

\begin{table}[t]
    \centering
\begin{tabular}{lccccc}
\toprule
method &       DML &  DML Iterative &   Lasso &  OLS Linear &  OLS Poly \\
\midrule
True Value               &      2.96 &           2.96 &    2.96 &        2.96 &      2.96 \\
Average Derivative       &     2.958 &           2.94 &   2.683 &       3.246 &     2.939 \\
Bias                     & -0.002252 &       -0.02015 & -0.2771 &      0.2861 &   -0.0208 \\
Standard Deviation       &    0.2991 &          0.341 &  0.3581 &      0.3311 &    0.5573 \\
MSE $\tau$               &   0.08013 &         0.1069 &  0.1957 &      0.1836 &    0.3009 \\
Coverage (95\% C.I.)                &     0.924 &          0.962 &   0.224 &       0.886 &      0.950 \\
MSE $\gamma$ In Sample   &     1.951 &          1.951 &   1.951 &       2.338 &     1.515 \\
MSE $\gamma$ Cross Folds &     2.048 &          2.048 &   2.048 &       2.454 &     10.04 \\
\bottomrule
\end{tabular}

    \caption{Summary of derivative estimates from 1000 bootstrap trials of our simulation procedure. Bias is the average of the estimated value of the derivative minus the true value of the derivative in each simulation draw. ``MSE $\tau$'' is the mean squared error between the true average derivative and the estimated average derivative in each simulation draw, while ``MSE  $\gamma$ in sample'' and ``MSE $\gamma$ cross-folds'' refer to the mean squared error of regression from own-sample and out-of-sample estimation. }
    \label{tab:simulation results}
\end{table}

Both DML and DML Iterative procedures results in a de-biased estimate of the average derivative. 
This is especially clear relative to Lasso without any correction. The bias of Lasso is 100 times larger than the bias of our proposed method.  
Our DML estimates also have substantially lower bias than OLS using the untransformed covariates (OLS Linear) and lower bias than OLS using the basis function transformation (OLS Poly). 
These results are consistent with expectations, as misspecification of OLS Linear can induce a bias in estimating the average derivative, and OLS Poly may be overfitting the data (as shown by the high error in MSE cross-folds).
Our optimization-based DML approach results in 10 times lower bias than DML Iterative, but does lose some coverage.  

We also compare the mean squared error of estimating the true parameter (MSE $\tau$), as this incorporates both  bias and variance of estimating $\hat{\tau}$. 
The MSE $\tau$ using DML is the lowest among all models considered, roughly half the magnitude from Lasso or OLS Linear and one quarter the magnitude from OLS Poly. 
The MSE $\tau$ for DML is also bit lower than that from using DML Iterative, which has the closest performance to our estimator. 
This shows that our optimization solution leads to not only a lower bias in estimating $\tau$, but also a lower variance, though we are losing a bit of coverage.  

These results demonstrate the value of DML and DML Iterative for applied research. Our methods provide an alternative to OLS that preserves the low standard errors of OLS Linear and the low bias of flexible modeling such as OLS Poly. Due to the regularization of Lasso, our estimates have considerably lower standard errors than OLS Poly even though they use the same set of basis functions. 
Figure \ref{fig:ridge plot of sims } visualizes this benefit: the error from DML and OLS Poly are both approximately centered around 0, but the distribution of errors using OLS Poly is much wider. 

In Appendix \ref{sec::appendix simulations}, we include results from simulation exercises with different specifications.
We change the number of time periods and the number of covariates. 
In all simulation trials, the MSE $\tau$ of our DML estimator is the lowest among all estimators considered.

\section{Application} \label{sec::application}
\label{sec:application}

We now apply our estimator to study the effect of extreme heat on U.S. corn yields, a central question in the climate–agriculture literature. Specifically, we estimate the elasticity of yields with respect to exposure to extreme heat, following the framework of \citet{Burke2016}. We compare estimates from their baseline specification—a low-dimensional linear regression—to those obtained from our more flexible double machine-learning estimator.

To illustrate the economic significance of our estimates, we translate them into projected damages from extreme heat in 2050 under median climate scenarios. We also estimate the dose–response curve, which traces how the marginal effect of extreme heat varies with counties' baseline exposure levels.

\subsection{Data Description}

We draw on county-level agricultural reports from the USDA Survey of Agriculture, combined with weather data from the GridMET dataset \citep{abatzoglou2013development}. Our sample closely follows \citet{Burke2016}, focusing on U.S. counties east of the 100\mydegree West meridian,\footnote{This region is commonly studied in U.S. agriculture due to the heavy reliance on subsidized irrigation west of this meridian.} but differs in two respects. First, we extend the coverage to 1980–2019. Second, we restrict to a balanced panel, maintaining the same set of counties throughout the analysis. Using this dataset, we replicate the linear specification of \citet{Burke2016} and recover results closely aligned with their findings.



\subsection{Empirical Results}

We begin by replicating the analysis of \citet{Burke2016},\footnote{Specifically, we estimate the specification reported in their Table 1, column 7.} which we present in Equation \eqref{eq:be_lm_mod}. The outcome variable $Y_{i,t}$ is the log of corn yields in county $i$ and year $t$, measured in bushels per acre. The key treatment variable is $GDD_{i,t;29:\infty}$, defined as the total number of growing degree days (GDD) above 29\mydegree C during the March–August growing season. Additional regressors include $GDD_{i,t;0:29}$, the number of growing degree days below 29\mydegree C, and precipitation variables $Prec_{i,t;p<50}$ and $Prec_{i,t;p\geq50}$, which measure total county-level precipitation (in centimeters) below and above 50 cm, respectively.
\begin{equation}
\label{eq:be_lm_mod}
\Delta Y_{i,t}  =  \beta_1 \Delta GDD_{it; 0:29} + \beta_2 \Delta GDD_{it; 29: \infty} + \beta_3 \Delta Prec_{it; p < 50} + \beta_4 \Delta Prec_{it; p > 50} +  \Delta \varepsilon_{i,t}. 
\end{equation}

The treatment effect of interest in \citet{Burke2016} is the APD which in their linear model is
$\beta_2$. 
In our replication of their linear model, we obtain an estimate of the ADP of $-0.0046$, which is nearly identical to the original estimate of $-0.0048$ reported by \citet{Burke2016}. 
These replication results are shown in Table \ref{tab:ag results table}, column 1.

Next, we extend the \citet{Burke2016} specification by incorporating additional weather controls available in our dataset. These include minimum and maximum relative humidity, mean specific humidity, mean wind speed, and mean downward shortwave radiation at the surface (a measure of solar radiation). Adding these covariates to the linear model yields an APD estimate of $-0.0047$, reported in Table \ref{tab:ag results table}, column 2.

Next, we apply our DML estimator using only the four original variables from \citet{Burke2016} 
($GDD_{i,t;0:29}$, $\Delta GDD_{i,t;29:\infty}$, $\Delta Prec_{i,t;p<50}$, $\Delta Prec_{i,t;p>50}$). 
We allow for fifth-order polynomials of these variables as well as their interactions. 
As shown in Table \ref{tab:ag results table}, column 3, the estimated APD is 
$-0.0055$, 
approximately 30 percent larger in magnitude than the estimate of $-0.0043$ reported in \citet{Burke2016}. 
When we expand the DML specification to include the additional weather variables described above, 
the effect becomes $-0.0064$ (Table \ref{tab:ag results table}, column 4), nearly 50 percent larger 
than the original \citet{Burke2016} estimate.
\begin{table}[h]
\small  
\centering 
\begin{tabular}{lcccc}
\toprule
Method &  B\&E &   B\&E + Covar &      DML &         DML + Covar \\
\midrule
Average Derivative       &   -0.0043 &   -0.0047 &  -0.0055 &   -0.0064 \\
                         &  (0.0001) &  (0.0001) &  (0.0001) &  (0.0001) \\
Number of Observations   &       22308 &       22308 &       22308 &       22308 \\
Number of Covariates     &           4 &          9 &          25 &          68 \\
\bottomrule
\end{tabular}
\caption{\footnotesize Estimates of elasticity of corn yields with respect to increase in growing season exposure to extreme heat. B\&E denotes the \citet{Burke2016} specification.
}
\label{tab:ag results table}
\end{table}

The average derivative estimates can be interpreted as the elasticity of corn yields with respect to additional exposure to extreme heat, holding all other weather variables constant. 
That is, each estimate is the percentage by which yields change with an additional growing degree day of heat exposure above 29\mydegree C over the growing season. 
The magnitude is relatively large. In our DML + Covar estimate, this suggests that an increase of a single growing degree day above 29\mydegree C is associated with corn yields declining by 0.64\%.

\begin{figure}[h]
     \centering
         \centering
         \includegraphics[scale=.65]{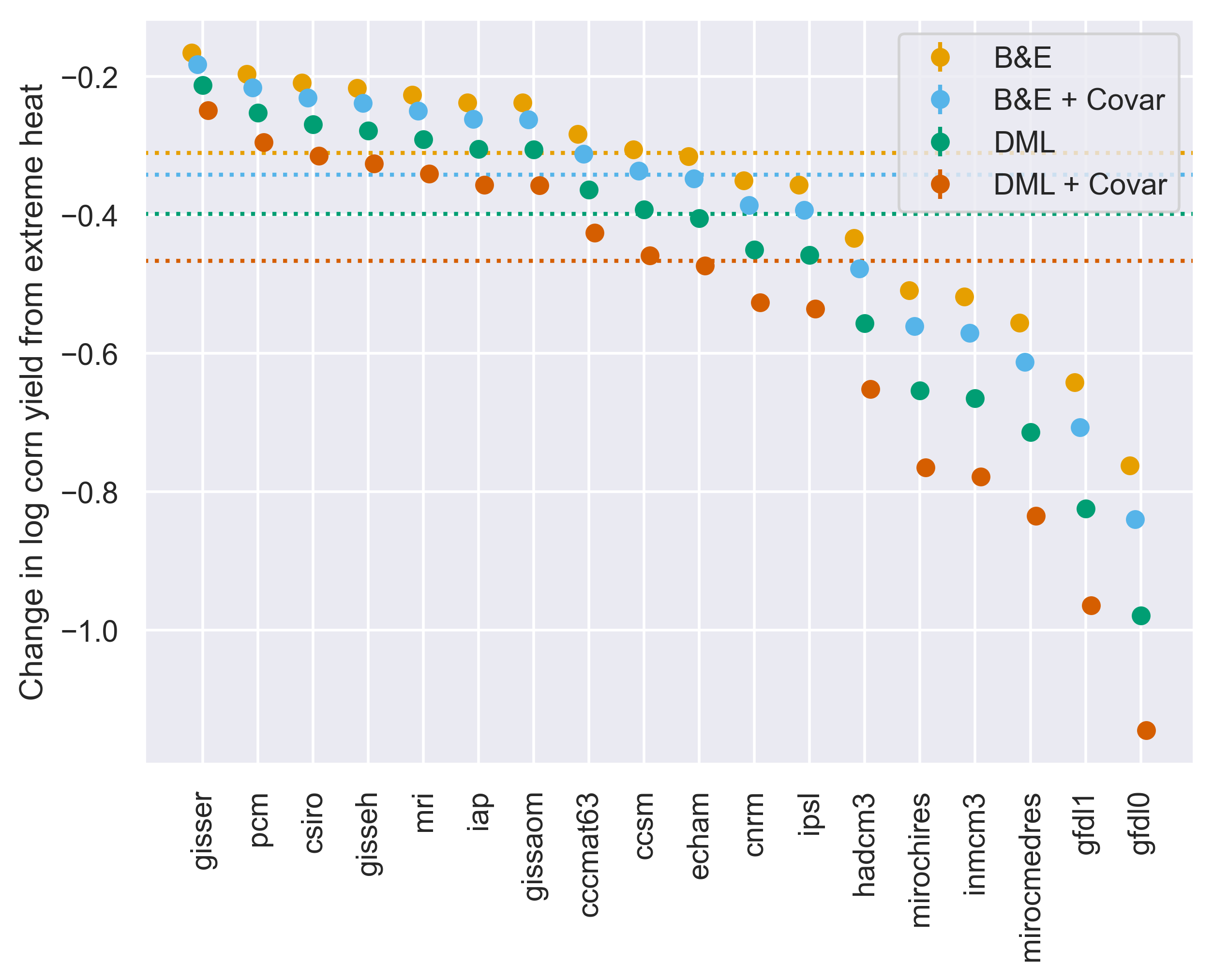}
        \caption{ \footnotesize The x-axis gives the 18 global climate models, y-axis gives the extrapolated impacts of extreme heat to crop yields by the year 2050. 
        Each dot represents a central estimate from a model, and the error bar represents the 95\% confidence interval. 
        Dotted line represents the median value across climate models.
        We follow \citet{Burke2016} and weight each observation by the area of corn planted in that county in that year.
        \label{fig:climate extrapolation}
}
\end{figure}

To gauge the economic significance of the four elasticities reported in Table \ref{tab:ag results table}, we project the overall change in corn yields by 2050 attributable solely to increased exposure to extreme heat, following the approach of \citet{Burke2016}. For each elasticity, we combine our estimates with climate projections from 18 global climate models under the A1B emissions scenario. Specifically, we multiply 
each elasticity by the weighted average increase in damaging heat between 2015 and 2050, where weights 
correspond to average county-level corn acreage. The county-level warming projections are generously 
provided by \citet{Burke2016}. Figure \ref{fig:climate extrapolation} presents the projected impacts for each of the four elasticities across the climate models.

The projected damages in Figure \eqref{fig:climate extrapolation} are economically substantial. In the median emissions scenario, log yields are $0.310$ to $0.466$ lower than a in world that does not experience climate change. To put this in dollar value, the 2017 Census of Agriculture reported the total value of sales of corn in the United States as \$51.2 billion. 
That range of damage estimates translates to a dollar value of \$13.7-\$19.1 billion (in 2017 dollars), or 26-37 percent of total yields lost.
Our DML elasticity, shown in green in Figure \eqref{fig:climate extrapolation}, leads to an estimated damage of \$16.8 billion, or 33\% total yields lost.
 
In Figure \eqref{fig:climate extrapolation} we also see that the projections differ significantly by which elasticity is used. The difference between the B\&E (yellow) and the DML (green) projection is significant with a p-value less than 0.001. Median damage estimates using DML model instead of the B\&E model correspond to an additional \$3.17 billion  per year in extreme heat-related damages. This difference is based completely on the difference in the models, rather than on different variables. The difference is even larger at \$5.4 billion if we use the DML + covar elasticity.  

Although projections vary substantially across climate models, Figure \ref{fig:climate extrapolation} illustrates the economic significance of using our estimator. It is important to note that these elasticities capture only the direct effect of extreme heat and do not account for potential adaptation to climate change. Moreover, each elasticity reflects the marginal effect at the observed distribution of weather conditions; as climate change alters this distribution, the implied projections will also evolve.

\begin{figure}[h]
\centering
{
\captionsetup{justification=centering}
   \includegraphics[scale = .5]{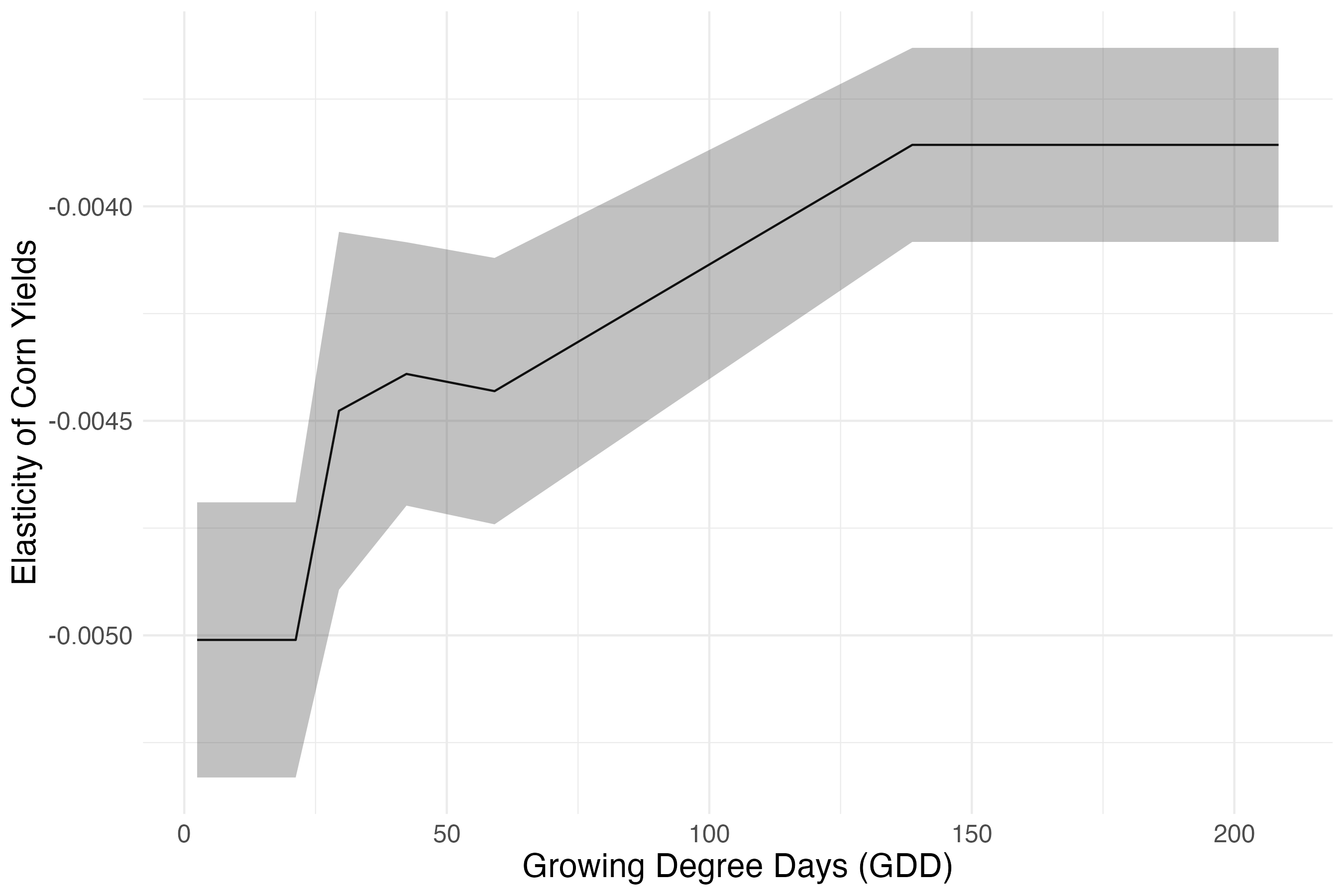}

     }
        \caption[Coarse dose response]{ \footnotesize Dose response function of the elasticity of corn yields with respect to extreme heat exposure, for quintiles of the extreme heat distribution. Error bars show the 95\% confidence interval. }
        \label{fig:dose response}
\end{figure}

Finally, we estimate a coarse dose–response function to assess how the impact of extreme heat varies 
with baseline exposure. Counties are divided into quintiles based on their average level of damaging 
heat, and our estimation procedure is applied separately within each group. The analysis uses the full 1980–2019 sample and the set of variables considered by \citet{Burke2016}. The resulting dose–response 
estimates are presented in Figure \ref{fig:dose response}.

The parameter estimates are strictly negative and statistically significant at the $p = 0.001$ level across all quintiles, indicating that extreme heat reduces yields even in counties with high baseline exposure. This finding is consistent with \citet{Schlenker2009}, who document persistent damages from extreme heat in southern U.S. regions where such temperatures are relatively common. The magnitude of the effect is largest in the lowest quintile of exposure and smallest in the highest, suggesting that damages attenuate somewhat in counties with greater historical exposure.

\section{Conclusion} \label{sec::conclusion}

Continuous treatments such as prices, pollution, and weather shocks are central to applied economics, yet standard linear two-way fixed effects models can be badly misspecified when the true relationships are nonlinear. This paper develops an ADML estimator that addresses this challenge by combining unrestricted additive fixed effects with flexible machine-learning methods. Importantly, it accommodates rich heterogeneity in treatment effects, capturing nonlinearities, interactions, and variation across units, time, and the dose–response curve. Our estimator is consistent for the population average partial derivative, asymptotically normal, and incorporates computational refinements that reduce bias and remove numerical approximation error.

Through simulations, we show that the estimator achieves substantially lower bias and variance compared to both linear specifications and high-order polynomial or standard ML approaches. Applying the method to U.S. corn yields under extreme heat, we find elasticities roughly 50 percent larger than piecewise-linear benchmarks, with no loss of precision. This implies materially higher projected damages from climate change, underscoring the economic importance of using flexible methods for continuous treatments.

More broadly, the framework provides applied researchers with a practical tool for analyzing continuous treatments in panel data without sacrificing the credibility of additive fixed effects. We expect the approach to be useful in a wide range of settings where nonlinear relationships are central, from environmental and health economics to taxation and trade.

\newpage

\printbibliography
\newpage
\appendix
\section{Asymptotic Normality}
\label{section:asymptotic_normality}

Proving that our estimator is asymptotically normal follows from Theorem 14 of \citet{chernozhukov2016locally}. Theorem 14 gives us that if Assumptions \eqref{assump:mild_mean_square}, \eqref{assump:interaction_rate}, and \eqref{assump:double_robust} below are satisfied, asymptotic normality follows. In this appendix we explain either why our estimator satisfies these assumptions, or give primitive conditions that justify the assumptions. 

We start with notation used in the three assumptions before introducing them. Recall from equation \eqref{eqn:average_derivative} from earlier our parameter of interest $ \tau_0$ is defined to be the solution to the following moment equation.

\begin{equation*}
    \begin{aligned}
    \tau_0 &=  \E[ m(W, \Delta \gamma) ] = \E\left[  \frac{\partial \Delta \gamma_{0}(D_{i,t}, X_{i,t}, D_{i,t-1}, X_{i,t-1})}{\partial D_{i,t}} \right]. 
\end{aligned}
\end{equation*}

We define 

\begin{equation}
    g(W, \Delta \gamma, \tau) = m(W, \Delta \gamma) - \tau.
\end{equation}

We define $\phi$ to be our first step influence function. In our application $\phi = \alpha_0(Y - \gamma_{0,t})$. 
Adding the first step influence function to our moment function to create our de-biased moment function $\psi$.  
\begin{align}
        \psi(W, \Delta \gamma, \alpha_0, \tau_0) &=  g(W, \Delta \gamma, \tau) + \phi(W, \Delta \gamma, \alpha_0, \tau_0). \\
        &=   g(W, \Delta \gamma, \tau) + \alpha_0(\Delta Y - \Delta \gamma_{0}).
\end{align}

We also define  $\hat{\delta}_{\ell}$ \footnote{note in the original paper by \citet{chernozhukov2016locally} this object was defined to be capital $\hat{\Delta}_{\ell}$ but given that capital delta is used to describe differences over time we changed notation } below. This is known as the interaction reminder, it shows up in the decomposition of $\hat{\psi} - \psi$ , and we have to make an assumption about its behavior for asymptotic linearity. 

\begin{equation}
\begin{aligned}
        \hat{\delta}_{\ell} &= 
    \hat{\alpha}_{\ell}(\Delta Y - \Delta \hat{\gamma}_{\ell}) - \hat{\alpha}_{\ell}(\Delta Y - \Delta{\gamma}_{0}) - \alpha_0(\Delta Y - \Delta \hat{\gamma}_{\ell}) - \alpha_0(\Delta Y - \Delta {\gamma}_{0}). \\
    &= (\hat{\alpha}_{\ell} - \alpha_0)(\Delta \hat{\gamma}_{\ell} - \Delta \gamma_{0}).
\end{aligned}
\end{equation}

 We use the notation $\| \cdot \|_{F,2}$ to denote the $L_2(F)$-norm and $\| \cdot \|$ for the $\ell^2$ vector norm. 

\begin{assumption} (mild mean-square consistency) 
\label{assump:mild_mean_square}: $E[\| \psi(W, \tau_0, \Delta \gamma_{0}, \alpha_0)^2 \|] < \infty$

\begin{itemize}
    \item[i)] $\int \| g(w, \Delta \hat{\gamma_{\ell}}, \tau_0) - \Delta g(w, \gamma_{0}, \tau_0) \|^2 F_0(dw) \xrightarrow[]{p} 0$.
    \item[ii)]
    
    $\int \| \alpha_0(Y - \Delta \hat{\gamma}_{\ell}) - \alpha_0(Y - \Delta {\gamma}_{0}) \|^2 F_0(dw) \xrightarrow[]{p} 0$.

    \item[iii)] 
     $\int \| \hat{\alpha}_{\ell}(\Delta Y - \Delta {\gamma}_{0}) - \alpha_0(\Delta Y - \Delta {\gamma}_{0}) \|^2 F_0(dw) \xrightarrow[]{p} 0$.

\end{itemize}
\end{assumption}

These are mild mean-square consistency conditions for $\Delta \hat{\gamma_{\ell}}$ and $(\hat{\alpha_{\ell}}, \Tilde{\theta}_{\ell} )$.

\begin{assumption}
\label{assump:interaction_rate} (rate on interaction remainder) For each $\ell = 1, \cdots, L$ 

\begin{equation}
    \sqrt{n} \int  \hat{\delta}_{\ell} (w) F_0(d_w)  \xrightarrow[]{p} 0, \int \| \hat{\delta}_{\ell} \|^2 (w) F_0(d_w)  \xrightarrow[]{p} 0.
\end{equation}

\end{assumption}

\begin{assumption}

\label{assump:double_robust} (double robust) For each $\ell = 1, \cdots, L$ 

\begin{itemize}
    \item[i)] $\int \phi(w, \Delta \gamma_{0}, \hat{\alpha_0}, \theta_0 ) F_0(dw) = 0$ with probability approaching one
    \item[ii)] $\E[{\psi}(\Delta \gamma, \alpha_0, \theta_0)]$ is affine in $\gamma$
\end{itemize}

\end{assumption}

As noted in \citet{chernozhukov2016locally}, in our case because  $\E[{\psi}(\gamma, \alpha_0, \theta_0)]$ is affine in $\gamma$, Assumption 3 imposes no conditions additional to Assumption 1 and 2, but we write it for clarity.

\subsection{Primitive Conditions for Assumptions}

\subsubsection{Primitive Conditions for Assumption \eqref{assump:mild_mean_square}}

Now we go into detail about the primitive conditions for the assumptions starting with Assumption \ref{assump:mild_mean_square} part i). Plugging in the definition of $g$ into the equation, we get that this assumption is the following. Given that our moment is one dimensional, we drop the norm notation for clarity. 

\begin{equation}
    \int \bigg( (\frac{\partial \Delta \hat{\gamma_{\ell}}}{\partial D_{i,t}} - \theta_0)  - (\frac{\partial \Delta {\gamma_{0}}}{\partial D_{i,t}} - \theta_0)  \bigg)^2 F_0(dw) \xrightarrow[]{p} 0.
\end{equation}

\begin{equation}
    \int \bigg( \frac{\partial \Delta \hat{\gamma_{\ell}}}{\partial D_{i,t}}   - \frac{\partial \Delta {\gamma_{0}}}{\partial D_{i,t}}   \bigg)^2 F_0(dw) \xrightarrow[]{p} 0.
\end{equation}

For conciseness, we write this with expectation notation.  

\begin{equation}
\label{eq:mean_sq_consis_exp_note}
     \E \bigg[ \bigg(  \frac{\partial \Delta \hat{\gamma_{\ell}}}{\partial D_{i,t}}   - \frac{\partial \Delta {\gamma_{0}}}{\partial D_{i,t}}   \bigg)^2 \bigg] \xrightarrow[]{p} 0.
\end{equation}

This is a mean square consistency condition for the average derivative. Mean square consistency for the average derivative follows from two primitive assumptions. First, that we have mean square consistency in the first step,  as stated in Assumption \eqref{assump:mean_square_consistency_first_step}. Second, that we have mean square continuity for average derivative, conditions for which are given in Assumption \eqref{assump:primitive_mean_square_continuity_average_derivative}. The proof showing that the conditions in Assumption \eqref{assump:primitive_mean_square_continuity_average_derivative} are sufficient for mean square continuity for the average derivative  proven is given in Lemma \eqref{lemma:Poincare}.

We start by writing these two assumptions and show how, together, they imply mean-square consistency for the average derivative.   

The first primitive assumption is mean square consistency of the first step estimator $\hat{\gamma}$. 

\begin{assumption} (mean square consistency first step)
\label{assump:mean_square_consistency_first_step}
\begin{equation}
    \| \Delta \hat{\gamma_{\ell}} - \Delta \gamma_{0} \|_{F,2} \xrightarrow[]{p} 0.
\end{equation}

\end{assumption}

The second primitive assumption is mean square continuity for the average derivative. In Assumption \ref{assump:primitive_mean_square_continuity_average_derivative} we give conditions for mean square continuity to hold for our average derivative. We prove that these conditions are sufficient in Lemma \eqref{lemma:Poincare}, where we prove a weak reverse Poincare inequality argument as in \citet{chernozhukov2021simple} - this time for the difference of two functions. 

\begin{assumption}
\label{assump:primitive_mean_square_continuity_average_derivative} Primitive conditions for Lemma \eqref{lemma:Poincare}

Assume that $f(d_t|d_{t-1}, x_t, x_{t-1} )$ vanishes for each $d_t$ in the boundary of the support of $D$ given $D_{t-1} = d_{t-1}$, $X_t = x_t$, $X_{t-1} = x_{t-1}$ almost everywhere. Next, assume the following restrictions on our function class $\Gamma \subset \mathbf{L}_2$.

\begin{itemize}
   
    \item[i)] $\Gamma$ consists of functions $\Delta \gamma$ that are twice continuously differentiable 

     \item[ii)] For each $\Delta \gamma$ in  $\Gamma$, $\|\kappa_{\gamma}\|_{pr,2} < \infty$

     where 

     $ \kappa_{\Delta \gamma} = \{- \partial_d \log f(d_t | x_t, d_{t-1}, x_{t-1}) \}\{\partial_d \gamma(D_t, X_t, D_{t-1}, X_{t-1}) \} - \partial_d^2 \Delta \gamma(D_t, X_t, D_{t-1}, X_{t-1})$

     We have that $\sup_{\Delta \gamma \in \Gamma} \|\kappa_{\gamma}\|_{pr,2} < \infty$ if either the following conditions hold

    \item $ \| \partial_d \log  f(d_{t}|d_{t-1}, x_t, x_{t-1}) \|_{pr,2} < \infty$ and for all $\Delta \gamma$ in  $\Gamma$,  $\| \partial_d \Delta \gamma \|_{\infty} < \infty$ and $\| \partial_d^2 \Delta \gamma \|_{\infty} < \infty$
    
    \item      $-\partial_d \log  f(d_{t}|d_{t-1}, x_t, x_{t-1})$ is bounded above and for all $\Delta \gamma \in \Gamma$, $\| \partial_d \Delta \gamma \|_{pr,2} < \infty$ and $\| \partial_d^2 \Delta \gamma \|_{pr,2} < \infty$
\end{itemize}

\end{assumption}

We need these assumptions on the conditional density of the treatment variable for the same reason we need the classic overlap condition in causal inference. The importance of the overlap conditions in the cross-sectional continuous treatment case is discussed in \cite{klosin2021automatic}. The important difference in the panel context is that the overlap condition is not just in terms of covariates in the current time period, but also in terms of last time periods treatment and covariates. That is, we require variation overtime in treatment and covariates in order for the overlap condition to hold.

\begin{lemma} (A weak reverse Poincare inequality for function differences)
\label{lemma:Poincare}. Assume that the conditions of Assumption \ref{assump:primitive_mean_square_continuity_average_derivative} hold, then  

\begin{equation}
    \E{(\frac{\partial \Delta \gamma}{\partial D_{i,t}}^2)} \leq \|\kappa_{\gamma}\|_{pr,2} [\E(\Delta \gamma^2)]^{\frac{1}{2}}
\end{equation}

\end{lemma}

\begin{proof}
\small

\begin{align}
     &\E[\{\partial_d \Delta \gamma_{0}(D_{i,t}, X_{i,t}, D_{i,t-1}, X_{i,t-1})\}^2] \\
     &= \int \{\partial_d \Delta \gamma_{0}(D_{i,t}, X_{i,t}, D_{i,t-1}, X_{i,t-1})\}^2 {f}(d_t,x_t, d_{t-1}, x_{t-1}) d_{d_t} d_{x_t} d_{d_{t-1}} d_{x_{t-1}} \\
    &= \int \{\partial_d \Delta \gamma_{0}(D_{i,t}, X_{i,t}, D_{i,t-1}, X_{i,t-1})\}^2 {f}(d_t|x_t, d_{t-1}, x_{t-1}) f(x_t, d_{t-1}, x_{t-1}) d_{d_t} d_{x_t} d_{d_{t-1}} d_{x_{t-1}} \\
&=  \int \{\partial_d \Delta \gamma_{0}(D_{i,t}, X_{i,t}, D_{i,t-1}, X_{i,t-1})\} {f}(d_t|x_t, d_{t-1}, x_{t-1}) \{\partial_d \Delta \gamma_{0}(D_{i,t}, X_{i,t}, D_{i,t-1}, X_{i,t-1})\} \\
&\quad \quad \times f(x_t, d_{t-1}, x_{t-1}) d_{d_t} d_{x_t} d_{d_{t-1}} d_{x_{t-1}} \\
&= - \int  \Delta \gamma_{0}(D_{i,t}, X_{i,t}, D_{i,t-1}, X_{i,t-1}) \partial_d \{{f}(d_t|x_t, d_{t-1}, x_{t-1})  \Delta \gamma_{0}(D_{i,t}, X_{i,t}, D_{i,t-1}, X_{i,t-1})\} \\
&\quad \quad \times f(x_t, d_{t-1}, x_{t-1}) d_{d_t} d_{x_t} d_{d_{t-1}} d_{x_{t-1}} \\
          &=   \int \partial_d \Delta \gamma(D_{i,t}, X_{i,t}, D_{i,t-1}, X_{i,t-1}) \kappa_{\Delta \gamma}  f(d_t|x_{t}, d_{t-1}, x_{t-1}) f(x_t, d_{t-1}, x_{t-1}) d_{d_t} d_{x_t} d_{d_{t-1}} d_{x_{t-1}} \\
          &\leq \|\Delta \gamma \| \|\kappa_{\Delta \gamma}\|
\end{align}    
\end{proof}

\begin{lemma}(mean square consistency for the average derivative)
Assumption \eqref{assump:mean_square_consistency_first_step} and Lemma \eqref{lemma:Poincare} imply mean square consistency for the average derivative.

\end{lemma}

\begin{proof}
    \begin{align}
           \E \bigg[ \bigg(  \frac{\partial \Delta \hat{ \gamma_{\ell}}}{\partial D_{i,t}}   - \frac{\partial \Delta {\gamma_{0}}}{\partial D_{i,t}}   \bigg)^2 \bigg] &= \E \bigg[ \bigg(  \frac{\partial (\Delta \hat{\gamma_{\ell}} - \Delta \gamma_{0})}{\partial D_{i,t}}    \bigg)^2  \bigg] \\
           &\leq  \|\kappa_{\gamma}\|_{pr,2} [\E{( \Delta \hat{\gamma}_{\ell} - \Delta \gamma_{0})^2}]^{\frac{1}{2}}
           \\
           &= \|\kappa_{\gamma}\|_{pr,2}  \| \Delta \hat{\gamma_{\ell}} - \Delta \gamma_{0} \|_{F,2} \xrightarrow[]{p} 0
    \end{align}
\end{proof}

We can bound the left hand side of equation \eqref{eq:mean_sq_consis_exp_note} using the two assumptions. First the linearity of the average derivative gives us the first equality below. Then the inequality follows by mean square continuity since it holds for every realization of  $\Delta \hat{\gamma}$.  Then the last line follows by Lemma \eqref{lemma:Poincare}. Hence we have mean square consistency condition for the average derivative.

\subsubsection{Primitive Conditions for Assumption \eqref{assump:interaction_rate}}
\label{sec:Primitive ConditionsforAssumption2}

The rate condition on the interaction term of Assumption \eqref{assump:interaction_rate} requires that $\hat{\alpha}_{\ell}$ and $\hat{\gamma}_{\ell}$ can be estimatedfast enough. 

\begin{equation}
\label{eq:interaction}
    \| \hat{\alpha}_{\ell} - \alpha_0 \|_{F,2} \|\Delta \hat{\gamma}_{\ell} - \Delta \gamma_{0} \|_{F,2} = o_p(n^{-1/2})
\end{equation}

The rates and conditions for the Lasso estimation of $\hat{\alpha}_{\ell}$ and $\Delta \hat{\gamma}_{\ell}$  to satisfy \eqref{eq:interaction} are discussed in detail in \citet{chernozhukov2022automatic}. We provide a brief overview here, often taking language about intuition from \citet{klosin2021automatic}.

\begin{assumption} (Bounded dictionary) 
\label{assump:bounded_basis}
There exists a $C$ such that with probability one 
\begin{equation}
    \max_{1 \leq j \leq p} |b_j| \leq C
\end{equation}
\end{assumption}

The next two assumptions - Assumptions \ref{assump:alpha_slow_rate} and \ref{assump:alpha_fast_rate} control the complexity of the true function $\alpha_0$. Intuitively, the less complex $\alpha_0$, the faster we can estimate it. Therefore the complexity of $\alpha_0$ governs the convergence rate $\|\hat{\alpha}_0 - \hat{\alpha}\|$. When using the estimator we can assume either Assumption \ref{assump:alpha_slow_rate} or \ref{assump:alpha_fast_rate} - whichever we find plausible. If we use Assumption  \ref{assump:alpha_slow_rate} we will get a slower rate, and if we use Assumption \ref{assump:alpha_fast_rate} we will get a faster rate. We call the complexity in Assumption \ref{assump:alpha_slow_rate} the dense regime and in Assumption \ref{assump:alpha_fast_rate} the sparse regime.  

\begin{assumption} 
\label{assump:alpha_slow_rate} (Dense regime)
For every $n$, there exists some $p \times 1$  sequence of coefficients $\rho_n$ and a positive constant $C < \infty$ such that $|\rho_n|_{1} \leq C$ and $\|\alpha_0 - b' \rho_n \|^2 = O(\sqrt{\frac{\ln(p)}{n}})$
\end{assumption}

Assumption \ref{assump:alpha_slow_rate} says that $\alpha_0$ can be approximated by our dictionary $b$. An example of when this assumption holds is when  $\alpha_0$ is a linear combination of the elements of the $b$ dictionary.

\begin{assumption} 
\label{assump:alpha_fast_rate} (Sparse regime)
Assume that the following hold.
\begin{enumerate}
    \item There exists $C, \xi >0$ such that for all $\bar{s}$ with  $\bar{s} \leq C(\sqrt{\frac{\ln(p)}{n}})^{\frac{-2}{(1 + 2 \xi)}}$  there is a $\bar{\rho} \in \R^p$ with $|\bar{\rho}|_1$ and $\bar{s}$ nonzero elements s.t. 
    
    \begin{equation}
        \| \alpha_0 - b' \bar{\rho} \| \leq C (\bar{s})^{- \xi}
    \end{equation}
    
    \item $Q = \E[ b b']$ is nonsingular and has the largest eigenvalue uniformly bounded in $n$ 
    
    \item  for $\rho=\bar{\rho}$  and $\rho=\arg\min_{\rho}\{\left\Vert \alpha_{0}-b^{\prime}\bar{\rho}\right\Vert
^{2}+2r_{L}\sum_{j=1}^{p}\left\vert \rho_{j}\right\vert \}$\textit{
there is }$k>3$ such such that 

\begin{equation}
\label{eq:restricted_eigenvalue_condition}
    \inf_{\{\delta:\delta\neq0,\sum_{j\in\mathcal{J}_{\rho_{}}^{c}}|\delta
_{j}|\leq k\sum_{j\in\mathcal{J}_{\rho_{}}}|\delta_{j}|\}}\frac
{\delta^{\prime}Q\delta}{\sum_{j\in\mathcal{J}_{\rho_{}}}\delta_{j}^{2}}>0
\end{equation}

where $\mathcal{J}_{\rho_{}} = support(p)$

\end{enumerate}
\end{assumption}

Part 3 of assumption \ref{assump:alpha_fast_rate} is a population version of the restricted eigenvalue condition of \citet{bickel2009simultaneous} as adapted in \citet{chernozhukov2022automatic}. A clear introduction of the restricted eigenvalue condition is given in \citet{tibshirani2016closer}.

\begin{assumption}(Regularization) \label{assump:regularization}
\begin{equation}
    r_{n} = a_n (\sqrt{\frac{\ln(p)}{n}}) \text{ for some } a_n \rightarrow \infty
\end{equation}
\end{assumption}

To satisfy this assumption we set $a_n = \ln(\ln(n))$, following \cite{chatterjee2015prediction}. In practice, we pick a data-driven following \cite{chernozhukov2022automatic}.

\begin{assumption} 
\label{assump:regressionrate}(Regression rate)

For each  $\ell=1,...,L \text{ and for any } t \in \mathcal{T}$  

\begin{equation}
    \| \Delta \hat{\gamma} - \Delta \gamma_0\| = O_p(n^{-d_{\gamma}})
\end{equation}

\begin{enumerate}
    \item in the dense regime, $d_{\gamma} \in (\frac{1}{4}, \frac{1}{2})$ 
    \item in the sparse regime, $d_\gamma \in (\frac{1}{ 2} - \frac{ \xi}{1 + 2 \xi}, \frac{1}{2})$
\end{enumerate}
\end{assumption}

These regime-specific bounds on $d_{\gamma}$ are sufficient condition to controls the interaction remainder.

\subsection{Cluster-Robust Variance}
We use a common assumption in panel data that observations are arbitrarily correlated within a panel unit but uncorrelated between panel units. 
Our cluster-robust form of the asymptotic variance follows from Theorem 3 of \citet{chernozhukov2022automatic} and \citet{chernozhukov2019demand}.
The cluster-robust form for the asymptotic variance:
\begin{equation}
\label{eqn:cluster_robust}
    \hat{V} = \frac{1}{n (T - 1)} \sum_{\ell=1}^L \sum_{i \in I_L}  \sum_{t =2}^T\sum_{t' =2}^T\hat{\psi}_{\ell; it}\hat{\psi}_{\ell; it'} 
\end{equation}
where $\hat{\psi}_{\ell; it} := m(\hat{\gamma}, W_{i,t}) + \alpha(W_{i,t}) (\Delta y_{i,t} - \Delta \hat{\gamma}(D_{i,t}, X_{i,t})) - \hat{\tau}$. 

This form comes because we are estimating the variance of the average derivative, allowing for arbitrary correlation within panel units. 
We illustrate this by computing the variance of the average effect:
\begin{equation}
    \var(\hat{\tau}) =  \var \left( \frac{1}{n (T -1)}\sum_{i} \sum_{t} \hat{\tau}_{i,t} \right) = \left( \frac{1}{n (T -1)}\right)^2 \sum_{i} \sum_{t} \sum_{t' } \cov (\hat{\tau}_{i,t} , \hat{\tau}_{it'}) 
\end{equation}
Plugging in the sample form of the covariance, realizing that $\hat{\psi}_{\ell; it} = \hat{\tau}_{\ell; it} - \E[\hat{\tau}_{i,t}]$, and multiplying by $n (T - 1)$ gives the estimate of the asymptotic variance. 

That is, 
\begin{equation*}
     \hat{V} \xrightarrow{p} V
\end{equation*}
where 
\begin{equation*}
    \frac{1}{\sqrt{nT}} (\hat{\tau} - \tau_0) \xrightarrow{d} \mathcal{N}(0,V) 
\end{equation*}

In our  procedure, we use a weighted form of the cluster-robust covariance estimate. 
Let $w_{i,t}$ be the per-unit weight, and let $\Bar{w} := \sum_{i \in \mathcal{I}}\sum_t^{T - 1} w_{i,t}$ 
Then, the variance of the weighted average effect is: 
\begin{equation}
    \var(\hat{\tau}) =  \var \left( \frac{1}{\Bar{w} }\sum_{i} \sum_{t} w_{i,t} \hat{\tau}_{i,t} \right) =  \frac{1}{\Bar{w} ^2} \sum_{i} \sum_{t} \sum_{t' } w_{i,t} w_{it'} \cov (\hat{\tau}_{i,t} , \hat{\tau}_{it'}) 
\end{equation}

\section{Details of Estimation Procedure} \label{sec:details of estimation}

\subsection{Analytical vs. Numerical Derivative}
\label{sec:numerical differentiation}
 We propose calculating the derivative analytically, instead of using numerical methods.
In current DML and ADML papers, derivatives are computed using numerical differentiation.
To explain what we mean by analytical vs numerical, let's consider a function $\gamma(D,X)$, and let's say we have an estimate of this function  $\hat{\gamma}(D,X)$ and we want to estimate the derivative at point $D  = D_0$ and $X = X_0$. To estimate the derivative numerically, we could use Newton's difference quotient (also known as a first-order divided difference) and pick some small $h$. 
    
\begin{equation}
\label{eq:numerical_derivative}
    \frac{\partial \hat{\gamma}(D_0, X_0)}{\partial D} = \lim_{h \rightarrow 0}  \frac{ \hat{\gamma}(D_0 + h, X_0) -  \hat{\gamma}(D_0, X_0)}{ h} 
\end{equation}

Or an alternative, using a symmetric difference: 
\begin{equation}
\label{eq:symmetric_difference}
    \frac{\partial \hat{\gamma}(D_0, X_0)}{\partial D} = \lim_{h \rightarrow 0}  \frac{ \hat{\gamma}(D_0 + h, X_0) -  \hat{\gamma}(D_0 - h, X_0)}{ 2h} 
\end{equation}

It is well known that this procedure can introduce biases, either through formula error or rounding error. 
Formula error is introduced because, for most cases, the difference between the true derivative and the numerical approximation is decreasing in $h$. Formula error is most relevant when $h$ is large. 
Rounding error is introduced during computation with small $h$ as computers must round floating-point numbers in order to carry out the computation.

There are approaches for reducing the error of numerical differentiation although these approaches are still under development, especially for noisy data. 
Simple modifications include taking a symmetric difference instead of a one-sided difference. 
The problem of numerical differentiation with noisy data is more  challenging and the subject of ongoing research (\cite{mboup2009numerical,chartrand2011numerical,chartrand2017numerical,van2020numerical}).
By taking analytical derivatives of our basis function, we avoid these numerical challenges.

\subsection{Tuning }
\label{sec:tuning alpha and gamma}

In this section we explain tuning $\hat{\alpha}$ and $\hat{\gamma}$.

We use a data-driven process to select the hyperparameters for Lasso and the Riesz representer function. 
We select hyperparameters by minimizing loss on the test set during each fold of our cross-validation procedure. 
Let $\hat{\gamma}_{\ell}$ and $\hat{\alpha}_{\ell}$ denote the estimates of $\gamma$ and $\alpha$ trained using indices not in set $\ell$ using the above procedures, for the given hyperparameter value. Let $\hat{\beta}_\ell$ and $\hat{\rho}_\ell$ be the corresponding estimates of parameter vectors in our sparse linear models. 
Recall that $W_{\ell}$ denotes all observations in the fold $\ell$. 
Let $\mathcal{L}_\gamma(\gamma, W_{\ell}; r_L)$ be the sum of squared error of the function $\gamma$ with the hyperparameter $r_L$ on the data in $\mathcal{I}_{\ell}$:
\begin{equation}
\mathcal{L}_\gamma(\gamma, W_{\ell}; r_L) = \sum_{i}\sum_{t} (\Delta \hat{\gamma}_\ell(D_{i,t}, X_{i,t}) - \Delta Y_{i,t})^2
\end{equation}
Let $\mathcal{L}_\alpha(\alpha, \mathcal{I}_{\ell}; r_\alpha)$ be the loss function of the function $\alpha$ with the hyperparameter $r_\alpha$ on the data in $\mathcal{I}_{\ell}$. 
That loss function is the sum of the distance between $\alpha_0$ and $\hat{\alpha}$ using our dictionary of basis functions. See Appendix \ref{sec::origin alpha} for an explanation of this loss function. Minimizing this distance is equivalent to minimizing: 
\begin{equation}
\mathcal{L}_\alpha(\hat{\alpha}_{\ell}, W_{\ell}; r_\alpha) = \sum_{i}\sum_{t} -2 b_D(D_{i,t}, X_{i,t})' \hat{\rho}_{\ell} + \hat{\rho}_{\ell}' \Delta b(D_{i,t}, X_{i,t})\Delta b(D_{i,t}, X_{i,t})'\hat{\rho}_{\ell}
\end{equation}

Then, select hyperparameter $r_L$ that minimize test-set mean squared error of the regression: 
\begin{equation}
    r_L = \argmin_{r} \frac{1}{n (T - 1)} \sum_{\ell = 1}^k \mathcal{L}_\gamma(\hat{\gamma}_\ell, W_{\ell}; r)
\end{equation}
And select hyparameter $r_\alpha$ that minimizes test-set loss of the Riesz representer: 
\begin{equation}
    r_\alpha = \argmin_{r}  \frac{1}{n (T - 1)}\sum_{\ell = 1}^k \mathcal{L}_\alpha(\hat{\alpha}_{\ell}, W_{\ell}; r)
\end{equation}

In our preferred estimation procedure, we use a weighted version of these minimization problems. 
Let $w_{i,t}$ be the per-observation weight, and $\Bar{w} := \sum_{i \in \mathcal{I}} \sum_t^{T - 1}$ be the sum of weights. 
Then, the weighted form of the loss functions and minimization problems is: 
\begin{equation}
\mathcal{L}_\gamma(\gamma, W_{\ell}; r_L) = \sum_{i}\sum_{t} w_{i,t} (\Delta \hat{\gamma}_\ell(D_{i,t}, X_{i,t}) - \Delta Y_{i,t})^2
\end{equation}
\begin{equation}
    r_L = \argmin_{r} \frac{1}{\Bar{w}} \sum_{\ell = 1}^k \mathcal{L}_\gamma(\hat{\gamma}_\ell, W_{\ell}; r)
\end{equation}
\begin{equation}
\mathcal{L}_\alpha(\hat{\alpha}_{\ell}, W_{\ell}; r_\alpha) = \sum_{i}\sum_{t} w_{i,t} \left\{-2 b_D(D_{i,t}, X_{i,t})' \hat{\rho}_{\ell} + \hat{\rho}_{\ell}' \Delta b(D_{i,t}, X_{i,t})\Delta b(D_{i,t}, X_{i,t})'\hat{\rho}_{\ell}\right\}
\end{equation}
\begin{equation}
    r_\alpha = \argmin_{r}  \frac{1}{\Bar{w}}\sum_{\ell = 1}^k \mathcal{L}_\alpha(\hat{\alpha}_{\ell}, W_{\ell}; r)
\end{equation}

\subsection{Normalization} \label{sec:normalization}

Many ML models perform better when the independent variables are standardized, that is when the data has mean zero and variance 1. 
In this section, we include some details about how to conduct this standardization so that the researcher is able to recover derivatives after that step. 

For each basis function $b^j$ for $j = 1, \dots, p$ in the dictionary $b$, define the mean and standard deviation of the transformed data: $\displaystyle \mu^j := \E[b^j(W_i)]$ and $\displaystyle \sigma^j := \sqrt{\E[(b^j(W_i) - \mu^j)^2]}$.
Their sample equivalents are:
$\displaystyle \hat{\mu}^j := \frac{1}{n}\sum_{i} b^j(W_i)$ and $\displaystyle \hat{\sigma}^j := \sqrt{\frac{1}{N - 1}\sum_{i} (b^j(W_i) - \hat{\mu}^j)^2}$.

To generate a standardized basis function, we apply the following transformation: $\dot{b}^j := (b^j(W_i) - \hat{\mu}^j) / \hat{\sigma}^j$. 
We assumed that there was some $\beta_0$ such that $\gamma_{0,t}(W_i) = \beta_0 b(W_i)$.
Then there exists $\tilde{\beta}_0$ such that $\gamma_{0,t}(W_i) = \tilde{\beta}_0 \dot{b}(W_i) + C$, where $\tilde{\beta}_0^j \sigma^j = \beta_0^j$ for all $j$ components, $\dot{b}$ is the dictionary of all standardized basis functions $\dot{b}^j$, and $C$ is some generic constant. 
This is easy to confirm via algebraic manipulation. 
When we take differences to remove the unobserved individual fixed effect, this $C$ term is also removed. 

Recall that the average derivative is $\E[\beta_0 b_D(W_i)]$.
We could also write this in terms of $\tilde{\beta}_0$, with the fact that $\beta_0^j = \tilde{\beta}_0^j \sigma^j$.
Let $\sigma = \{\sigma^1, \dots, \sigma^p\}$, and $\sigma^{-1} = \{1 / \sigma^1, \dots,  1/\sigma^p\}$
Then, we write this relationship more compactly as $\E[(\tilde{\beta}_0 \circ \sigma^{-1}) b_D(W_i)$, where $\circ$ is elementwise multiplication or the Hadamard product (i.e. $\tilde{\beta}_0 \circ \sigma^{-1} = \{\tilde{\beta}_0^1 / \sigma^1, \dots, \tilde{\beta}_0^p / \sigma^p \}$).

In our estimation procedure, we found it more convenient to produce scaled derivatives of each basis function and multiply them by the beta estimate from scaled data.
\footnote{This form can also be motivated by the chain rule, taking the derivative of the standardized data. In this case, an additional bias correction would be necessary because we are estimating $\hat{\mu}^j$ and $\hat{\sigma}^j$. See Appendix \ref{sec:normalization} for more details.}
Define the scaled derivative of each basis function, $\dot{b}_D^j := b_D^j / \hat{\sigma}^j$. 
Then, the average derivative is: $\E[\tilde{\beta}_0 \dot{b}_D^j(W_{i,t})]$. 
These procedures are equivalent, as can be confirmed through algebraic manipulation. 

This suggests our procedure to standardize data and recover the derivative:
\begin{enumerate}
    \item For each basis function $b^j$ for $j = 1, \dots, p$, find $\hat{\mu}^j $ and $\hat{\sigma}^j$. Store these estimates. 
    \item Create the sample standardized basis function and its derivative, $\dot{b}^j := (b^j(W_i) - \hat{\mu}^j) / \hat{\sigma}^j$ and $\dot{b}_D^j := b_D^j / \hat{\sigma}^j$.
    \item Find an estimate $\hat{\tilde{\beta}}$ that satisfies the regression $\E[\Delta y_{i,t}] = \hat{\tilde{\beta}} \Delta \dot{b}(W_{i,t})$. 
    
    Where $\Delta \dot{b}(W_{i,t}) := \dot{b}(W_{i,t}) - \dot{b}(W_{i,t-1})$. This could be via OLS or a cross-folds Lasso procedure. 
    \item Estimate the average derivative as $\E[\hat{\tilde{\beta}} \dot{b}_D(W_{i,t})]$.
\end{enumerate}

Standardization of our basis functions is also relevant for estimating the Riesz representer. 
After the standardization, our Riesz representer now takes the form $\hat{\alpha}(W_{i,t}) = \dot{b}(W_{i,t}) \hat{\rho}$.
Before standardizing the data, we had an estimator of the form: 
\begin{equation}
    \hat{\rho}_\text{original} = \argmin_\rho -2\hat{M}\rho + \rho'\hat{Q}\rho + \lambda |\rho|_1
\end{equation}
where $\hat{M} =\E[ b_D(W_{i,t}) ]$ and $\hat{Q} = \E[\Delta b(W_{i,t})' \Delta b(W_{i,t})]$.
After applying the standardization, we are taking the derivative of the standardized basis functions, so this estimator now takes the form: 
\begin{equation}
\label{eq:standardized riesz representer}
\hat{\rho} = \argmin_\rho -2\hat{\dot{M}}\rho + \rho'\hat{\dot{Q}}\rho + \lambda |\rho|_1
\end{equation}
where $\hat{\dot{M}} =\E[ \dot{b}_D(W_{i,t}) ]$ and $\hat{\dot{Q}} = \E[\Delta \dot{b}(W_{i,t})' \Delta \dot{b}(W_{i,t})]$.
We use $\dot{b}^j_D(W_{i,t}) = b^j_D(W_{i,t}) / \hat{\sigma}^j$, an approximation to the true derivative of the standardized basis function. 

Note that the true derivative differs from this expression because standardization involves estimating the mean and variance of the dataset, each of which depends on the treatment variable. 
Below, we show that the difference between this approximation and the true value are negligible for large $N$. 
More precisely, $b^j_D(W_{i,t}) / \hat{\sigma}^j - \dot{b}^j_D(W_{i,t}) = O(1/N)$.
The difference between the approximation and the true derivative of the normalized basis function is asymptotically negligible, as it converges faster than the $1/\sqrt{N}$ rate.

In the remainder of this section, we derive this fact.
Keeping the convention that $\xi_D$ denotes the partial derivative of $\xi$ with respect to $D$, we can write the partial derivative of the normalized basis function:
\[
\frac{\partial }{\partial D_i} \frac{b^j(W_i) - \hat{\mu}^j}{\hat{\sigma}^j} = \frac{(b_D^j(W_i) - \hat{\mu}^j_D) \hat{\sigma}^j - (b^j(W_i) - \mu^j) \hat{\sigma}^j_D}{(\hat{\sigma}^{j})^{2}}
\]
We then expand the derivatives of the estimates of mean and standard deviation. Note that this expansion introduces another index, as we are summing over all observations to compute the partial derivative with respect to the treatment variable $D_i$. We label this second index $k$: 
\begin{flalign*}
&(b_D^j(W_i) - \hat{\mu}^j_D) = b_D^j(W_i) - 1 / N b_D^j(W_i) = \frac{N-1}{N} b_D^j(W_i) \\
&\hat{\sigma}^j_D = \frac{1}{2}\left(\frac{\sum_{k} (b^j(W_k) - \hat{\mu}^j)^2}{N - 1}\right)^{-1/2} \frac{1}{N - 1} \sum_{k} 2 (b^j(W_k) - \hat{\mu}^j)(b_D^j(W_k) - \hat{\mu_D}^j) 
\end{flalign*}
The expression for the standard deviation can be simplified somewhat; note that because the partial derivative is 0 for all $k \neq i$
\[
\sum_{k}  (b^j(W_k) - \hat{\mu}^j)(b_D^j(W_k) - \hat{\mu_D}^j) = (b^j(W_i) - \hat{\mu}^j)b_D^j(W_i) - \sum_{k} (b^j(W_k) - \hat{\mu}^j)\hat{\mu_D}^j 
\]
and because $\hat{\mu}^j$ is the mean of $b^j(W_k)$:
\[
\hat{\mu_D}^j \sum_{k} (b^j(W_k) - \hat{\mu}^j) = 0
\]

Plugging in these terms and rearranging into the above expression, we have: 
\[
\frac{\partial }{\partial D_i} \frac{b^j(W_i) - \hat{\mu}^j}{\hat{\sigma}^j} = \frac{b_D^j(W_i)}{\hat{\sigma}^j} \left( \frac{N - 1}{N} - \frac{(b^j(W_i) - \mu^j)^2}{(N - 1) (\hat{\sigma}^{j})^{2}} \right)
\]
Then we have: 
$\displaystyle \frac{b_D^j(W_i)}{\hat{\sigma}^j} - \frac{b_D^j(W_i)}{\hat{\sigma}^j} \left( \frac{N - 1}{N} - \frac{(b^j(W_i) - \mu^j)^2}{(N - 1) (\hat{\sigma}^{j})^{2}} \right) = O(1/N)$.
This establishes that the difference between our estimate of the derivative of the standardized basis function and its true derivative is asymptotically negligible.

\subsection{Riesz Representer Details} \label{sec::origin alpha}

Our goal is to find the estimator $\hat{\alpha}(W_{i,t})$ that minimizes the mean squared error (MSE) between $\hat{\alpha}$ and $\alpha_0$:
 \begin{equation}
     \hat{\alpha} = \argmin_\alpha \E[(\alpha_0(W_{i,t}) - \alpha(W_{i,t}))^2]
 \end{equation}
 
Plugging in our guess at the functional form, $\hat{\alpha} = \Delta b(W_{i,t}) \hat{\rho}$, we form the following regularized problem: 
\begin{align*}
\hat{\rho} &= \argmin_\rho \E\left[ (\alpha_0(W_{i,t}) - \Delta b(W_{i,t})\rho)^2\right] + \lambda |\rho|_1\\
&= \argmin_\rho \E[\alpha_0(W_{i,t})^2 - 2 \Delta b(W_{i,t}) \alpha_0(W_{i,t}) \rho + \rho ' \Delta b(W_{i,t})'\Delta b(W_{i,t}) \rho] +  \lambda |\rho|_1\\
&= \argmin_\rho -2\E[ b_D(W_{i,t}) ]\rho + \rho' \E[\Delta b(W_{i,t})' \Delta b(W_{i,t})]\rho +  \lambda |\rho|_1\\
&= \argmin_\rho -2\hat{M}\rho + \rho'\hat{Q}\rho + \lambda |\rho|_1
\end{align*}
where the 3rd equality comes from applying the Riesz Representation theorem, and the 4th equality comes from the definition $\hat{M} =\E[ b_D(W_{i,t}) ] $ and $\hat{Q} = \E[\Delta b(W_{i,t})' \Delta b(W_{i,t})]$. 

We use this estimator to find $\hat{\rho}$; we use an optimization package to find the optimal value of $\hat{\rho}$. \citet{chernozhukov2022automatic} provides an iterative approach. 

More generally, for any linear functional $m$ and any functional form for $\hat{\alpha}$:
\begin{align*}
\hat{\alpha} &= \argmin_\alpha \E[(\alpha_0(W_{i,t}) - \alpha(W_{i,t}))^2] \\
&= \argmin_\alpha \E[\alpha_0(W_{i,t})^2] - 2 \E[ \alpha_0(W_{i,t})\alpha(W_{i,t})] + \E[\alpha(W_{i,t})^2] \\
&= \argmin_\alpha -2\E[m(W_{i,t},\alpha)] + \E[\alpha(W_{i,t})^2]
\end{align*}
Plugging in our estimator form, we get $\E[m(W_{i,t},\alpha)] = \E[b_D(W_{i,t}) \rho]$ and $\E[\alpha(W_{i,t})^2] = \rho' \E[\Delta b(W_{i,t})' \Delta b(W_{i,t})]\rho$ , confirming the above result. 

In our preferred estimation procedure, we use a weighted version of this estimator. In this case, the expectation operator simply places unequal weight on observations as defined by our weighting scheme. 
That is, if $w_{i,t}$ is the per-unit weight and $\Bar{w} := \sum_{i \in \mathcal{I}} \sum_{t}^{T-1}$, $\E[X_{i,t}] = 1 / \Bar{w} \sum_{i \in \mathcal{I}} \sum_{t}^{T-1} w_{i,t} X_{i,t}$

\section{Heterogeneity Over Time}
\label{section:hetero_time}

This section helps build intuition for why we are able to estimate time-varying treatment effects in our first differences approach. We do this through giving two examples, first when there is homogeneous effects, and second when there is not.  

First, let us say that we have four time periods. Our DGP is 

\begin{equation}
    y_{i,t} = a_i + \tau D_{i,t} + \varepsilon_{i,t}
\end{equation}

Taking first differences does the following in order to control for the $a_i$ term. 

\begin{equation}
    (y_{i,t} - y_{i,t-1}) = \tau(D_{i,t} - D_{i, t-1}) + (\varepsilon_{i,t} - \varepsilon_{i,t-1})
\end{equation}

We can move the $\tau$ to the outside of the $(D_{i,t} - D_{i, t-1})$ difference because it is constant throughout time. To estimate $\tau$ we would have to create a new variable in the data $\Delta D_{i,t} = (D_{i,t} - D_{i, t-1})$ and the coefficient in front of it would be $\tau$. 

Now let us look at a second example, when there is heterogeneity in treatment over time. Let us say that $\tau_t$ is different in every time period. So there are $\tau_1$, $\tau_2$, $\tau_3$, $\tau_4$. We write our model in the following way. Here $1_t$ is an indicator variable for time period $t$.

\begin{equation}
    y_{i,t} = a_i + \sum_{k = 1}^4 \tau_k (D_{i,k} \times 1_{k =t}) + \varepsilon_{i,t}
\end{equation}

If we do first difference

\begin{equation}
    \begin{aligned}
    (y_{i,t} - y_{i,t-1}) &= \sum_{k = 1}^4 \tau_k (D_{i,k} \times 1_{k =t}) - \sum_{k = 1}^4 \tau_k (D_{i,k} \times 1_{k= t-1}) + (\varepsilon_{i,t} - \varepsilon_{i,t-1}) \\
    &=  \sum_{k = 1}^4 (\tau_k D_{i,k} \times (1_{k=t} - 1_{k=t-1})) + (\varepsilon_{i,t} - \varepsilon_{i,t-1})
\end{aligned}
\end{equation}

We are still able to recover the $\tau_t$ parameters by running the first difference regression, we just must work with the interaction of the difference of time indicator variables.  

\section{Additional Simulation Results}
\label{sec::appendix simulations}
Here, we include simulation results from different parameters of the data generating process described in Section \ref{sec::simulations}. 
Each table summarizes 1000 bootstrap trials, for the specified data generating process. We vary the number of covariates and the number of time periods in comparison to our main simulation results in the paper which have T = 2 and 20 covariates.  

These results show that, as expected, the bias generally decreases as the number of samples per covariate increases. This is particularly true of OLS Poly. 
OLS Poly has the lowest bias in these additional trials, but has higher mean squared error in estimating the true derivative (MSE $\tau$) than DML, DML Iterative, or Lasso in all trials. 
DML has lower bias than DML Iterative in Table \ref{tab:simulation results p10t2} and Table \ref{tab:simulation results p10t5}, and lower MSE $\tau$ in all trials. 

\begin{table}[ht!]
    \centering
\begin{tabular}{lccccc}
\toprule
method &       DML &  DML Iterative &   Lasso &  OLS Linear &  OLS Poly \\
\midrule
True Value               &     2.937 &          2.937 &   2.937 &       2.937 &     2.937 \\
Average Derivative       &     2.929 &          2.909 &   2.819 &       3.245 &     2.932 \\
Bias                     & -0.007124 &       -0.02712 & -0.1173 &      0.3082 & -0.004251 \\
Standard Deviation       &    0.3187 &         0.3498 &   0.311 &      0.3553 &    0.5017 \\
MSE $\tau$               &   0.09302 &         0.1141 &  0.1021 &      0.2152 &    0.2436 \\
Coverage                 &     0.906 &          0.942 &   0.287 &       0.845 &     0.942 \\
MSE $\gamma$ In Sample   &     1.945 &          1.945 &   1.945 &       2.349 &     1.752 \\
MSE $\gamma$ Cross Folds &     2.033 &          2.033 &   2.033 &       2.411 &      3.59 \\
\bottomrule
\end{tabular}

    \caption{Summary of derivative estimates from 1000 bootstrap trials of our simulation procedure. 
    Estimates use $N=1000$, $T = 2$, and $10$ covariates. Flexible basis functions include $3^\text{rd}$ order polynomial  functions of all terms and all interactions of $D$ and $X$ terms. After applying the basis function transformation, $p=124$. 
    Bias is the average of the estimated value of the derivative minus the true value of the derivative in each simulation draw. ``MSE $\tau$'' is the mean squared error between the true average derivative and the estimated average derivative in each simulation draw, while ``MSE  $\gamma$ in sample'' and ``MSE $\gamma$ cross-folds'' refer to the mean squared error of regression from own-sample and out-of-sample estimation. }
    \label{tab:simulation results p10t2}
\end{table}

\begin{table}[ht!]
    \centering
\begin{tabular}{lccccc}
\toprule
method &        DML &  DML Iterative &    Lasso &  OLS Linear &  OLS Poly \\
\midrule
True Value               &      2.936 &          2.936 &    2.936 &       2.936 &     2.936 \\
Average Derivative       &      2.935 &          2.922 &    2.864 &       3.239 &     2.936 \\
Bias                     & -0.0006326 &       -0.01354 & -0.07144 &      0.3038 & 0.0003044 \\
Standard Deviation       &     0.2258 &         0.2425 &   0.2187 &      0.2272 &    0.2925 \\
MSE $\tau$               &    0.03628 &        0.04374 &  0.03838 &      0.1293 &   0.06965 \\
Coverage                 &        0.9 &          0.923 &    0.289 &       0.678 &     0.937 \\
MSE $\gamma$ In Sample   &      1.988 &          1.988 &    1.988 &       2.366 &     1.918 \\
MSE $\gamma$ Cross Folds &       2.01 &           2.01 &     2.01 &       2.388 &     2.207 \\
\bottomrule
\end{tabular}

    \caption{Summary of derivative estimates from 1000 bootstrap trials of our simulation procedure. 
    Estimates use $N=1000$, $T = 5$, and $10$ covariates. Flexible basis functions include $3^\text{rd}$ order polynomial  functions of all terms and all interactions of $D$ and $X$ terms. After applying the basis function transformation, $p=124$. 
    Bias is the average of the estimated value of the derivative minus the true value of the derivative in each simulation draw. ``MSE $\tau$'' is the mean squared error between the true average derivative and the estimated average derivative in each simulation draw, while ``MSE  $\gamma$ in sample'' and ``MSE $\gamma$ cross-folds'' refer to the mean squared error of regression from own-sample and out-of-sample estimation. }
    \label{tab:simulation results p10t5}
\end{table}

\begin{table}[ht!]
    \centering
\begin{tabular}{lccccc}
\toprule
method &     DML &  DML Iterative &    Lasso &  OLS Linear &  OLS Poly \\
\midrule
True Value               &   2.964 &          2.964 &    2.964 &       2.964 &     2.964 \\
Average Derivative       &   2.977 &           2.96 &    2.902 &       3.258 &     2.973 \\
Bias                     & 0.01301 &      -0.003874 & -0.06199 &      0.2943 &  0.008642 \\
Standard Deviation       &  0.2292 &           0.25 &   0.2222 &      0.2351 &    0.2878 \\
MSE $\tau$               & 0.03879 &        0.04942 &   0.0391 &      0.1245 &   0.07208 \\
Coverage                 &  0.8747 &         0.9114 &     0.27 &      0.6976 &    0.9287 \\
MSE $\gamma$ In Sample   &    1.98 &           1.98 &     1.98 &       2.369 &     1.838 \\
MSE $\gamma$ Cross Folds &   2.013 &          2.013 &    2.013 &       2.409 &     2.577 \\
\bottomrule
\end{tabular}

    \caption{Summary of derivative estimates from 1000 bootstrap trials of our simulation procedure. 
    Estimates use $N=1000$, $T = 5$, and $20$ covariates. Flexible basis functions include $3^\text{rd}$ order polynomial  functions of all terms and all interactions of $D$ and $X$ terms. After applying the basis function transformation, $p=244$. 
    Bias is the average of the estimated value of the derivative minus the true value of the derivative in each simulation draw. ``MSE $\tau$'' is the mean squared error between the true average derivative and the estimated average derivative in each simulation draw, while ``MSE  $\gamma$ in sample'' and ``MSE $\gamma$ cross-folds'' refer to the mean squared error of regression from own-sample and out-of-sample estimation. }
    \label{tab:simulation results p20tt}
\end{table}

\newpage 

\end{document}